\documentclass[twocolumn,times,tighten,trackchanges]{aastex62}

\usepackage{amssymb,natbib,graphicx,subfigure,color,multirow}

\def\LCDM{$\Lambda\mbox{CDM}$}

\def\AREPO{{\small AREPO}}

\def\SUBFIND{{\small SUBFIND}}

\def\log{{\rm\thinspace log}}

\newcommand{\ifm}[1]{\relax\ifmmode#1\else$\mathsurround=0pt #1$\fi}
\newcommand{\be}{\begin{equation}}
\newcommand{\ee}{\end{equation}}
\newcommand{\bea}{\begin{eqnarray}}
\newcommand{\eea}{\end{eqnarray}}

\newcommand{\Fig}[1]{Figure (\ref{f:#1})}
\newcommand{\Figs}[2]{Figures (\ref{f:#1}) and (\ref{f:#2})}

\newcommand{\sggg}[1]{\textcolor{green}{[]}}

\def\dex{{\rm\thinspace dex}}

\def\pc{{\rm\thinspace pc}}
\def\kpc{{\rm\thinspace kpc}}

\def\Mpc{{\rm\thinspace Mpc}}

\newcommand{\hMpc}{\,\ifm{h^{-1}}{\rm Mpc}}
\newcommand{\hkpc}{\,\ifm{h^{-1}}{\rm kpc}}

\def\Msun{\hbox{$\rm\thinspace M_{\odot}$}}

\def\yr{{\rm\thinspace yr}}
\def\Myr{{\rm\thinspace Myr}}
\def\Gyr{{\rm\thinspace Gyr}}

\def\Msunyr{\Msun\yr^{-1}}
\def\Msunpc2{{\Msun\pc}^{-2}}
\def\Msunyrkpc2{{\Msun\yr^{-1}\kpc}^{-2}}

\def\magarcsec2{{\rm\thinspace mag\thinspace arcsec}^{-2}}

\shorttitle{The Butterfly Effect in Cosmological Simulations}
\shortauthors{Genel, S., et al.}

\begin{document}

\title{A Quantification of the Butterfly Effect in Cosmological Simulations and Implications for Galaxy Scaling Relations}

\author[0000-0002-3185-1540]{Shy Genel}
\affiliation{Center for Computational Astrophysics, Flatiron Institute, 162 Fifth Avenue, New York, NY 10010, USA}
\affiliation{Columbia Astrophysics Laboratory, Columbia University, 550 West 120th Street, New York, NY 10027, USA}
\author{Greg L.~Bryan}
\affiliation{Center for Computational Astrophysics, Flatiron Institute, 162 Fifth Avenue, New York, NY 10010, USA}
\affiliation{Department of Astronomy, Columbia University, 550 West 120th Street, New York, NY 10027, USA}
\author{Volker Springel}
\affiliation{Max-Planck-Institut f{\"u}r Astrophysik, Karl-Schwarzschild-Stra{\ss}e 1, 85740 Garching bei M{\"u}nchen, Germany}
\author{Lars Hernquist}
\affiliation{Institute for Theory and Computation, Harvard-Smithsonian Center for Astrophysics, 60 Garden Street, Cambridge, MA 02138, USA}
\author{Dylan Nelson}
\affiliation{Max-Planck-Institut f{\"u}r Astrophysik, Karl-Schwarzschild-Stra{\ss}e 1, 85740 Garching bei M{\"u}nchen, Germany}
\author{Annalisa Pillepich}
\affiliation{Max-Planck-Institut f{\"u}r Astronomie, K{\"o}nigstuhl 17, 69117 Heidelberg, Germany}
\author{Rainer Weinberger}
\affiliation{Heidelberg Institute for Theoretical Studies, Schloss-Wolfsbrunnenweg 35, 69118 Heidelberg, Germany}
\author{R{\"u}diger Pakmor}
\affiliation{Heidelberg Institute for Theoretical Studies, Schloss-Wolfsbrunnenweg 35, 69118 Heidelberg, Germany}
\author{Federico Marinacci}
\affiliation{Institute for Theory and Computation, Harvard-Smithsonian Center for Astrophysics, 60 Garden Street, Cambridge, MA 02138, USA}
\author{Mark Vogelsberger}
\affiliation{Department of Physics, Kavli Institute for Astrophysics and Space Research, MIT, Cambridge, MA 02139, USA}

\email{sgenel@flatironinstitute.org}

\begin{abstract}
We study the chaotic-like behavior of cosmological simulations by quantifying how minute perturbations grow over time and manifest as macroscopic differences in galaxy properties. When we run pairs of `shadow' simulations that are identical except for random minute initial displacements to particle positions (e.g.~of order $10^{-7}\pc$), the results diverge from each other at the individual galaxy level (while the statistical properties of the ensemble of galaxies are unchanged). After cosmological times, the global properties of pairs of `shadow' galaxies that are matched between the simulations differ from each other generally at a level of $\sim2-25\%$, depending on the considered physical quantity. We perform these experiments using cosmological volumes of $(25-50\Mpc/h)^3$ evolved either purely with dark matter, or with baryons and star-formation but no feedback, or using the full feedback model of the IllustrisTNG project. The runs cover four resolution levels spanning a factor of 512 in mass. We find that without feedback the differences between shadow galaxies generally become smaller as the resolution increases, but with the IllustrisTNG model the results are mostly converging toward a `floor'. This hints at the role of feedback in setting the chaotic properties of galaxy formation. Importantly, we compare the macroscopic differences between shadow galaxies to the overall scatter in various galaxy scaling relations, and conclude that for the star formation-mass and the Tully-Fisher relations the butterfly effect in our simulations contributes significantly to the overall scatter. We find that our results are robust to whether random numbers are used in the subgrid models or not. We discuss the implications for galaxy formation theory in general and for cosmological simulations in particular.
\end{abstract}

\keywords{galaxies: formation --- evolution --- cosmology: theory --- methods: numerical --- hydrodynamics --- chaos}

\section{Introduction}
\label{s:intro}

Cosmological simulations are the most general tool for theoretical studies of galaxy formation. Significant progress is continuously being made on their physical fidelity, numerical accuracy and computing power, and as a result, also on their realism. However, several factors hinder the prospect of accurately simulating our Universe. One well-known limitation stems from small scales: the need to model processes occurring on scales below the resolution of any given simulation using approximations (usually called `subgrid' models). Another limitation that is widely appreciated originates from large scales: our ignorance about the initial conditions of cosmological systems, whether our own Galaxy or our Universe as a whole (often referred to as `cosmic variance'). In this work we consider for the first time the possible consequences of a limitation that in some sense is a combination of the two: our ignorance about initial conditions on {\it small} rather than large scales.

The butterfly effect is the phenomenon whereby a dynamical system evolves in a macroscopically different manner due to a minute change in initial conditions. Systems that possess this property are often loosely referred to as chaotic. In this work we use the term `chaotic-like' to refer to phenomena related to the butterfly effect. A more formal definition of a chaotic system may involve the existence of a positive Lyapunov exponent, namely the exponential divergence of trajectories that are initially only infinitesimally separated. In regimes where we do identify an exponential growth of initially small differences, we refer to the timescale associated with this growth as the Lyapunov timescale, but in many cases the divergence we observe is not exponential and is therefore `chaotic-like'. Simulations that start from almost identical initial conditions are referred to here, following standard nomenclature in the context of chaos studies, as `shadow' simulations, and matched systems within these simulations, such as particles or galaxies, are also referred to as `shadow' versions of each other.

Chaotic-like systems can be found in diverse contexts in Astrophysics. Examples include the dynamics of planetary systems \citep{LaskarJ_89a}, N-body systems such as star clusters or dark matter halos \citep{HeggieD_91a,ElZantA_18a} as well as galactic disks and bars \citep{FuxR_01a,SellwoodJ_09a}, star-formation in turbulent molecular clouds \citep{AdamsF_04a,BateM_10a}, and the orbits of satellite galaxies, stellar streams and halo stars \citep{MaffioneN_15a,PriceWhelanA_16a,PriceWhelanA_16b}. Here we study the butterfly effect in a context that has hitherto been largely neglected: the galaxy formation process from cosmological initial conditions in the \LCDM{ }paradigm. To this end, we employ state-of-the-art cosmological hydrodynamical simulations and study the growth over cosmological timescales of minute perturbations applied to them. We also discuss the applicability of our results and conclusions beyond the realm of simulations, namely for the real universe.

Chaotic-like sensitivity to initial conditions in cosmological systems,
as a related yet distinct phenomenon from other discreteness effects (e.g.~\citealp{RomeoA_08a,vandenBoschF_18a}),
has been considered in a few cases before, dating back to \citet{SutoY_91}. For example, \citet{ThiebautJ_08a} measured the characteristic growth (Lyapunov) timescales of small differences between initial conditions in sets of otherwise identical cosmological pure N-body boxes. They found that chaos-like behavior appears on small, non-linear scales, but is absent on large, linear scales. Interestingly, some global properties of dark matter halos were found to be robust and stable to these magnified differences on the particle level, but not all. \citet{ThiebautJ_08a} identified several global halo properties that differed significantly between shadow versions of the same halos, such as spin and orientation of the velocity dispersion tensor.
\citet{ElZantA_18a} recently found that global properties of non-cosmological, equilibrium spherical N-body systems show an initial exponential growth of errors but then a saturation that converges toward zero as the number of particles is increased toward the collisionless limit. The direct relevance of this result to the case of halos developing from cosmological initial conditions is unknown and merits further research (see also \citealp{BenhaiemD_18a}).
\citet{KaurovA_18a} found that small-scale modifications to cosmological initial conditions propagate to much larger scales by the epoch of reionization, dramatically affecting simulation results such as the escape fraction.
In this paper we perform measurements that are similar in spirit to those of \citet{ThiebautJ_08a}, but on cosmological simulations that include baryons, hydrodynamics and galaxy formation models, and using different methods for introducing differences and measuring their growth.

Recently, while this paper was in preparation, \citet{KellerB_18a} investigated chaotic-like behavior seeded by roundoff errors in gravito-hydrodynamical simulations of a few individual galaxies, both from idealized and from cosmological initial conditions. With the codes they used, {\small GASOLINE2} \citep{WadsleyJ_17a} and {\small RAMSES} \citep{TeyssierR_02a}, repeated runs of the same setup resulted in different outcomes. They showed that the results of these different runs have normal distributions. In cases where the difference between two such shadow simulations grows to large values (even up to order unity), which often are associated with galaxy mergers, it tends to later converge back to the mean, a behavior they interpret as a result of negative feedback loops and global physical constraints on the system. They conclude that in order to determine the degree to which the results from simulations with different physical models truly differ from one another, the measured differences between them must be assessed keeping in mind the butterfly effect, namely with respect to differences that would occur merely by repeating runs with the same model.

In this work we quantify the differences between shadow hydrodynamical simulations of galaxies in the cosmological context. In contrast with \citet{KellerB_18a}, who have studied just a few individual galaxies and a large number of shadow simulations for each of them, we use `large-scale' (tens of $\Mpc$) cosmological boxes that contain thousands of galaxies, and use a small number of shadow simulations for each set of initial conditions. This allows us to quantify the average magnitude of the butterfly effect for a statistically representative galaxy population. In addition, we study how galaxies move due to the butterfly effect in parameter spaces combining several physical quantities that are related to each other through `scaling relations', and thereby quantify how much of the scatter in those relations is affected by the butterfly effect. The width, or scatter, in scaling relations is often considered to be no less fundamental than their shape parameters such as mean normalization and slope. For example, \citet{McGaughS_12a,McGaughS_15a} consider the very small scatter in the Tully-Fisher relation between galaxy luminosity and rotation speed \citep{TullyB_77a} as evidence toward modified gravity. The scatter around the mean relation between galaxy mass and star-formation rate (SFR) has also been studied extensively (e.g.~\citealp{TacchellaS_16a,MattheeJ_18a}), and it is believed to encode a variety of key processes in galaxy formation.

This paper is organized as follows. Section \ref{s:methods} describes the simulations we use and the analysis methods applied to them. Section \ref{s:results_individual} presents results for several individual galaxy properties from hydrodynamical cosmological simulations. Section \ref{s:results_relations} lays out the main results of this work, which concern several combinations of properties, namely scaling relations. Section \ref{s:summary} contains a summary and an extensive discussion. Finally, Appendix \ref{s:DMonly} briefly presents results from dark matter-only cosmological simulations, and Appendix \ref{s:verification} discusses several special sets of simulations run for numerical verification purposes.

\section{Methods}
\label{s:methods}

\subsection{Simulations}
\label{s:simulations}

\subsubsection{Code and Setup}
\label{s:simulations_setup}

We employ the MPI-parallel Tree-PM-moving-mesh code \AREPO{ }\citep{SpringelV_10a} to run three series of cosmological simulations, distinguished by different sets of physical components and models they include. Specifically, the DM-only series represents pure N-body simulations of cold dark matter; the No-feedback series adds baryons, hydrodynamics, radiative cooling, and star formation, utilizing the methods presented in \citet{VogelsbergerM_12a}; and the TNG series employs a more comprehensive treatment of the physics of galaxy formation, including in particular supermassive black holes as well as various feedback processes, utilizing the same models \citep{WeinbergerR_16a,PillepichA_16a} used for the IllustrisTNG project \citep{MarinacciF_17a,NaimanJ_17a,NelsonD_17a,PillepichA_17a,SpringelV_17a}. 

Each of these series is comprised of simulations at four resolution levels, the basic parameters of which are provided in Table \ref{t:resolution_levels}. The naming convention we use to distinguish the resolution levels is related to the spatial resolution. The $\epsilon=1$ resolution level, for example, is similar to (slightly worse than) the Illustris simulation \citep{GenelS_14a,VogelsbergerM_14a,VogelsbergerM_14b}, while the $\epsilon=0.5$ level has a mass resolution that is nearly five times better than Illustris. For the higher resolution levels we are limited by computational power to volumes of $(25\Mpc/h)^3$, but for the lower resolution levels we can afford to run larger volumes of $(50\Mpc/h)^3$, which is helpful for statistical power. The initial conditions for some of our cosmological boxes have been generated with N-GenIC \citep{SpringelV_05a} and are adopted from \citet{VogelsbergerM_13a}, and some generated with MUSIC \citep{HahnO_11a} especially for this study. We uniformly use a \LCDM{ }cosmology with $h=0.704$, $\sigma_8=0.809$, $n_s=0.963$, $\Omega_m=0.2726$, and (except for the DM-only series) $\Omega_b=0.0456$.

\begin{table*}
\begin{tabular*}{0.99\textwidth}{@{\extracolsep{\fill}}|c|c|c|c|c|c|}
\hline
resolution & dark matter gravitational                         & baryonic                    & dark matter                   & box size        & number of              \\
level      &   softening [comoving $\hkpc$]     & particle mass  [$h^{-1}\Msun$]    & particle mass  [$h^{-1}\Msun$]  & [$(\hMpc)^3$]  & dark matter particles  \\
\noalign{\vskip 0.5mm}
\hline
\hline
\noalign{\vskip 0.5mm}
$\epsilon=4$     & $4.0$                           & $9.4\times10^7$            & $4.7\times10^8$                 & $50^3$         & $256^3$    \\
$\epsilon=2$     & $2.0$                           & $1.2\times10^7$           & $5.9\times10^7$                & $50^3$         & $512^3$    \\
$\epsilon=1$     & $1.0$                           & $1.5\times10^6$           & $7.3\times10^6$                & $25^3$         & $512^3$    \\
$\epsilon=0.5$   & $0.5$                           & $1.8\times10^5$          & $9.2\times10^5$                & $25^3$         & $1024^3$   \\
\hline
\end{tabular*}
\caption{\small {\bf Properties of the different simulation resolution levels used in this study.} Our simulations are comprised of four resolution levels that span a factor of $8$ in spatial resolution and $512$ in mass resolution. Throughout the paper, they are referred to using the notation in the left-most column, based on their spatial resolution. In comparison to the IllustrisTNG simulations, the $\epsilon=1$ level is similar to (slightly worse than) the resolution of the TNG100, and so is the $\epsilon=2$ level with respect to TNG300. In addition to dark matter particles, whose number is provided in the right-most column, the initial conditions of hydrodynamical simulations include an identical number of gas cells.}
\label{t:resolution_levels}
\end{table*}

\subsubsection{Creating Shadow Simulations Using Minute Perturbations}
\label{s:simulations_shadows}

Each cosmological box is first evolved from its initial conditions at $z=127$ down to some final redshift, producing several snapshots at intermediate times. These snapshots are then used as initial conditions for what we call {\it sets of shadow simulations}, up to a unique minute perturbation that is applied to each of the shadow simulations in the set (described in the next paragraph). A set consisting of $N_s$ simulations contains then $N_s!(N_s-1)!/2$ pairs of shadow simulations, for which the setup and initial conditions are identical up to a minute perturbation. An overview of these sets is provided in Table \ref{t:simulations}, including the number of simulations and pairs in each set, as well as the perturbation (namely, initial) redshift and the final one. In most cases, unless otherwise noted, the shadow simulations produce snapshots at prescribed times starting $8\times10^5\yr$ after their initial time, namely the time the perturbations are introduced, in intervals increasing by a factor of two up to $4\times10^8\yr$ past the perturbation time\footnote{Since snapshots can only be written by our code at time steps when all particles are active, the snapshot times cannot be prescribed exactly, but are rounded to those special time steps. This implies that simulations with lower resolutions have a lesser ability to produce snapshots at very fine intervals. Accordingly, only the $\epsilon=0.5$ resolution level simulations can produce a snapshot as early as $8\times10^5\yr$ after their initial time, while for the $\epsilon=4$ resolution level the first snapshot is only written $5\times10^6\yr$ into the run. This can easily be changed by imposing a maximum time step, but that undesirably affects also the integration itself, as shown in Appendix \ref{s:verification} and is therefore not done in the main body of this work.}. This achieves high time resolution for following the early stages of the evolution of the perturbations. Thereafter, the snapshot separation is approximately equal in the logarithm of the cosmological scale factor. The total number of snapshots written by each shadow simulation between $z=5$ and $z=0$ is $\sim30$. In addition to the sets presented in Table \ref{t:simulations}, several special sets have been run for numerical verification reasons. These are described and discussed in Appendix \ref{s:verification}.

\begin{table*}
\begin{tabular}{|l|c|c|c|c|c|c|}
\hline
series & resolution level & number of sets (volumes) & number of simulations & resulting number of pairs & perturbation $z$ & final $z$ \\
\noalign{\vskip 0.5mm}
\hline
\hline
\noalign{\vskip 0.5mm}
\multirow{4}{*}{DM-only}
 & $\epsilon=4$ & 1 & 3 & 3 & 5 & 0 \\
 & $\epsilon=2$ & 1 & 3 & 3 & 5 & 0 \\
 & $\epsilon=1$ & 1 & 3 & 3 & 5 & 0 \\
 & $\epsilon=0.5$ & 1 & 2 & 1 & 5 & 0 \\
 \hline
\multirow{4}{*}{No-feedback}
 & $\epsilon=4$ & 1 & 3 & 3 & 5 & 0 \\
 & $\epsilon=2$ & 1 & 2 & 1 & 5 & 0 \\
 & $\epsilon=1$ & 1 & 3 & 3 & 5 & 0 \\
 & $\epsilon=0.5$ & 2 & $4+3$ & $6+3$ & 5 & 0.5 \\
 \hline
\multirow{4}{*}{TNG model}
 & $\epsilon=4$ & 1 & 3 & 3 & 5 & 0 \\
 & $\epsilon=2$ & 1 & 2 & 1 & 5 & 0 \\
 & $\epsilon=1$ & 1 & 3 & 3 & 5 & 0 \\
 & $\epsilon=0.5$ & 1 & 2 & 1 & 5 & 0 \\
\hline
\multirow{3}{1.8cm}{No-feedback; no random numbers}
 & $\epsilon=4$ & 1 & 2 & 1 & 5 & 0 \\
 & $\epsilon=2$ & 1 & 2 & 1 & 5 & 0 \\
 & $\epsilon=1$ & 1 & 2 & 1 & 5 & 0 \\
 \hline
\multirow{3}{1.8cm}{TNG model; no random numbers}
 & $\epsilon=4$ & 1 & 2 & 1 & 5 & 0 \\
 & $\epsilon=2$ & 1 & 2 & 1 & 5 & 0 \\
 & $\epsilon=1$ & 1 & 2 & 1 & 5 & 0 \\
\hline
\end{tabular}
\caption{\small {\bf An overview of the simulation suite used in this study.} We use three series of simulations (first column), each with a different physical model: simulations including only dark matter (DM-only), simulations with baryons and star-formation (No-feedback), and simulations with a full galaxy formation model (TNG model). In each series, there are four resolution levels (second column), most of which employ a single cosmological box, except for the high-resolution No-feedback case that uses two distinct boxes, providing two sets of shadow simulations (third column). The total number of simulations comprising each set is reported in the fourth column, and the resulting number of pairs of shadow simulations is reported in the fifth column (in the No-feedback $\epsilon=0.5$ case, the two numbers correspond to the two sets). The penultimate column reports the redshift at which the shadow simulations are perturbed and resumed, and the last one the final redshift to which they are evolved.}
\label{t:simulations}
\end{table*}

\begin{table*}
\begin{center}
\begin{tabular*}{0.912\textwidth}{|c|c|c|c|c|c|c|c|}
\hline
resolution & box size & \multicolumn{2}{c|}{$9<\log{M_*[h^{-1}\Msun]}<9.5$}   & \multicolumn{2}{c|}{$9.5<\log{M_*[h^{-1}\Msun]}<10$}  & \multicolumn{2}{c|}{$10<\log{M_*[h^{-1}\Msun]}<10.5$}  \\
\cline{3-8}
\cline{3-8}
level &  [$(\hMpc)^3$] & No-feedback &  TNG model & No-feedback &  TNG model  & No-feedback &  TNG model  \\
\hline
$\epsilon=4$     & $50^3$ & 781    & 596   & 1181  & 432 & 1455 & 366  \\
$\epsilon=2$     & $50^3$ & 5392   & 1201   & 5077  & 796 & 3377 & 637  \\
$\epsilon=1$     & $25^3$ & 2050    & 277    & 1399   & 168 & 653 & 108  \\
$\epsilon=0.5$   & $25^3$ & 3856    & 346    & 1899   & 214 & 688 & 133  \\
\hline
\end{tabular*}
\end{center}
\caption{\small {\bf Numbers of galaxies included in the analysis.} For each resolution level and for each of the two hydrodynamical models (without and with feedback) the number of galaxies in a single (arbitrarily selected) shadow simulation is given in three stellar mass bins. For each bin, the number of galaxies increases with box size and with better resolution, as well as when feedback is turned off. The intermediate mass bin corresponds to the one used in most figures throughout the paper. These numbers indicate the statistical power of our analysis by virtue of the large cosmological volumes employed.}
\label{t:galaxy_numbers}
\end{table*}

The `minute perturbation' applied to every simulation in every set is in most cases (unless noted otherwise) implemented as a displacement in the position of each and every particle in the snapshot that serves as the common initial conditions of the set. These displacements are applied only once, immediately after reading the snapshot data into memory and before any calculations are done to evolve the system. These displacements are applied in all three Cartesian spatial directions, and their magnitudes in each direction are $x_i r_i$, where $r_i$ the coordinate of the particle in the Cartesian direction $i$ and $x_i$ is drawn from a uniform distribution between (unless noted otherwise) $-5\times10^{-15}$ and $5\times10^{-15}$. Since particle positions are handled with double precision floating points, whose significand has a precision of 53 bits or $\approx16$ decimal digits, this range of possible displacements spanning $10^{-14}\times r_i$ translates into $\sim100$ possible values for the displacement of any given particle along each Cartesian axis.

With this design choice of limiting the displacements to a constant, small number of bits representing the position of each particle, the typical physical size of the displacement scales with the position in the box $r_i$. Given the box sizes we use, the maximal particle coordinates are of order tens of $\Mpc$, and the displacements are hence at most of order $10^{-7}\pc$ (comoving). An alternative possible design choice of keeping a constant physical displacement size across the box rather than a constant relative displacement size would be inconsequential to the results we present, for two reasons. First, due to the fact that for the vast majority of particles (except very close to the origin where all three $r_i$ are much smaller than the box size, or where all three $x_i$ happen to be $\ll1$) the magnitudes of the initial displacements are within the same order of magnitude.
Second, due to the fact that our results are largely insensitive to the magnitude of the initial perturbations, as demonstrated in Appendix \ref{s:verification_DM}.

\subsubsection{Discussion of Numerical Nuisance Parameters}
\label{s:simulations_nuisance}

Other than the application of a unique realization of displacements to each simulation, all shadow simulations in a given set are evolved identically, in terms of, e.g.~the Linux kernel, the executable\footnote{Compiled using \texttt{gcc} with the strong optimization configuration \texttt{-O3}, unless noted otherwise.}, the number of compute cores and MPI\footnote{Message Passing Interface.} tasks, the random number generator\footnote{Specifically, we employ the \texttt{gsl\_rng\_ranlxd1} random number generator from the GNU Scientific Library (\texttt{gsl-2.3}) with a seed of $42+r$, where $r$ denotes the MPI rank, unless noted otherwise.}, and so on\footnote{This does not include, however, the specific nodes on which the computation is done.}. We choose to directly introduce explicit perturbations so that we have full control over them. We could have, however, introduced them in a less explicit way by, for example, running each shadow simulation using a different number of MPI tasks. Such a choice would immediately introduce a different realization of round-off errors in the force calculation due to a different order of summation, generating a very similar outcome to perturbations we introduce `by hand' close to the machine precision level. The number of MPI tasks hence effectively serves as a nuisance parameter that modifies the results of a simulation through the arbitrary realization of round-off errors. Since any specific order of summation is arbitrary, no emergent sequence of round-off errors (and hence evolution of the system) is more correct than any other (whether and in what sense the ensemble of solutions to the system represents the true physical solution is a different question, see e.g.~\citealp{BoekholtT_15a,PortegiesZwartS_18a}). This is true even with the simplest set of physics, namely in pure N-body simulations, as well as in pure hydrodynamical simulations, let alone in a combined gravity and hydrodynamics case.

It is worth commenting, however, that when we use the exact same setup, keeping all factors described above fixed, and do not introduce any perturbation, namely running `the same simulation' more than once, our code produces results that are binary identical, remaining so even over integrations of billions of years of cosmic time\footnote{Note that the specific nodes on which the computations are done are {\it not} required to be kept fixed for the results of the calculations to be binary identical.}. This is achieved by a deterministic order of operations that is independent of machine noise such as communication speeds between different nodes, providing a deterministic emergent sequence of round-off errors. It is nevertheless important to realize that this feature of exact reproducibility has nothing to do with accuracy: the reproducible realization of round-off errors with a particular setup of our code is arbitrary, and is no more accurate than any other one. For example, the different arbitrary realization of round-off errors that our exact same code and setup would obtain if only the number of MPI tasks was modified is just as correct. 

In our simulations that on top of gravity and hydrodynamics include also star-formation there is an additional nuisance parameter that is worth discussing, which is the seed for the random number generator. Random numbers are used in our model in the star-formation and feedback process to determine where stars will form or galactic winds be launched \citep{SpringelV_03a}. This is necessary since the timescales associated with these processes are of order $\sim10\Myr-1\Gyr$, while simulation time steps can be as short as $0.1-1\Myr$. Therefore, star-forming gas cells have typically very low probabilities during individual time steps to be converted into stellar or `wind' particles. The realization of these probabilities into actual star-formation or wind-launching events is controlled by random numbers. With a fixed seed for the random number generator, two identical setups result in identical results. However, if the seed for the random number generator is modified, a different sequence of random numbers is generated, and stars will form at different times and positions. This will also be the case if the same seed, and hence random numbers sequence, is used but with a time or cell offset between two simulations. It is important to realize that differences in round-off errors, or the introduction of minute displacements as described above, will quickly develop into effective offsets in the random number sequence, and hence have the same effect. This is because once the round-off errors develop into a situation where the number of star-forming gas cells in one simulation is different from its shadow simulation, each individual cell will be affected by a modified series of random numbers.

In order to examine whether the usage of random numbers affects our results in any meaningful way, we run a few simulation sets that completely avoid them. For the reason explained in the previous paragraph, this necessarily implies that the subgrid physics model is modified as well. To remove the usage of random numbers, we change the subgrid model such that any gas cell that crosses the star-formation density threshold is immediately converted to a collisionless star particle. Similarly, in simulations with the TNG model, any such gas cell is converted into two collisionless particles, each with half of the original mass, one of which is a stellar particle and the other a wind particle. These modifications effectively change both the star-formation timescale and the wind mass loading factors in the model. More subtle changes are also applied to the directionality of both galactic winds and black hole feedback, such that they do not use random numbers. All these modifications result in galaxies that are physically different from those in the fiducial model, e.g.~in their gas contents and morphologies, but these differences are secondary to our purpose here. The important aspect is rather that the results become completely independent of the random number generator, and hence provide an important sanity check on our conclusions. The results of these tests are discussed in Section \ref{s:no_random}.

In Appendix \ref{s:verification} we present further tests in which the treatment of random numbers in our simulations is modified, and in particular examine how circumventing the effects of random numbers affects the early evolution of the differences between shadow simulations. In Section \ref{s:summary} we discuss the relation of the usage of random numbers in our models to the real universe. We refer the reader to these sections for further details.

\subsection{Analysis}
\label{s:analysis}

\subsubsection{Matching between Shadow Simulations}
\label{s:analysis_matching}

\begin{figure*}
\centering
\includegraphics[width=1.0\textwidth]{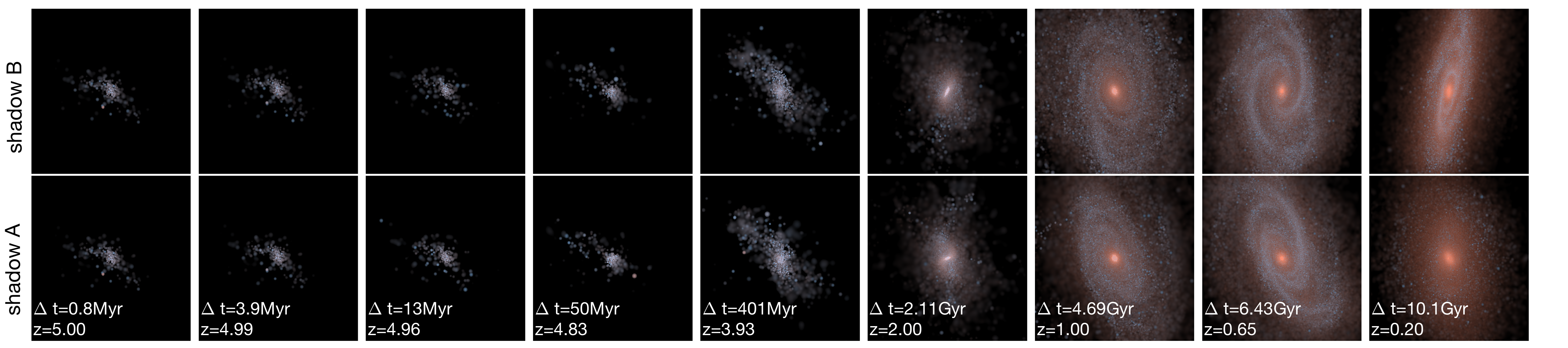}
\caption{A visual demonstration of the butterfly effect in the evolution of a pair of shadow galaxies. A galaxy in the initial ($z=5$) snapshot of the $\epsilon=0.5$ simulation set in our TNG-model series is followed over time (from left to right) in each of the two shadow simulations in the set (top/bottom rows). Each image is a color-composite representing the stellar luminosity in the (SDSS)r-g-(Johnson)B bands, and is centered at the most bound particle in the galaxy and projected along the z-axis of the simulation box. The redshift and time elapsed since the time a perturbation has been applied to the initial $z=5$ snapshot are indicated in the bottom row. Structural differences can be quite easily discerned at $\Delta t\sim2\Gyr$, but smaller differences, such as in the positions of individual stellar particles, can be seen as early as $\Delta t\sim4\Myr$. By $\Delta t\sim4-7\Gyr$, the initial perturbations have evolved into differences in the structure of the spiral arms and the overall orientation of the disk. At $z=0.2$, the galaxy has a prominent star-forming disk in the simulation shown in the top row, but in that shown in the bottom it has already largely quenched as a result of a gas ejection event by the central supermassive black hole, and hence has a markedly different color.}
\vspace{0.3cm}
\label{f:evolution_images}
\end{figure*}

The first analysis task given a set of shadow simulations is to match individual objects -- galaxies or dark matter halos -- between these simulations and thereby obtain a catalog of `shadow objects'. The type of objects that we match in practice is \SUBFIND{ }subhalos \citep{SpringelV_01}. These objects are matched between each pair of shadow simulations by identifying subhalos across simulations that have common dark matter particle IDs, namely according to commonalities of their Lagrangian patches. Specifically, the shadow subhalo in simulation B of a subhalo in simulation A is the subhalo in simulation B that contains the largest number of dark matter particles that are among the $N_p$ most bound dark matter particles in the subhalo in question from simulation A. The number $N_p$ is set to $1\%$ of the total number of dark matter particles in the subhalo in question, bounded by $20$ from below and $100$ from above. Further, if multiple halos from simulation A find the same match in simulation B, then only the most massive of them is kept as a valid match, and the rest are discarded\footnote{The determination of which simulation in a pair of shadow simulations is `A' and which is `B' is arbitrary. We also checked an alternative method: enforcing a bi-directional match by discarding all galaxies whose match's match is not themselves. This resulted in discarding $<5\%$ of the galaxies, and had virtually no effect on our results.}. We perform these matches for all subhalos with a stellar mass larger than $10^8h^{-1}\Msun$ in the hydrodynamical simulations or total mass larger than $10^{10}h^{-1}\Msun$ in the DM-only simulations. This procedure results typically in a matched fraction of $\sim98\%$. \Fig{evolution_images} presents mock stellar light images of a pair of matched shadow galaxies from a series of snapshots starting from shortly after the perturbation is applied and covering most of cosmic time.

In our analysis we narrow these matches down to include only those that are between two subhalos that are both the main subhalos of their Friends-Of-Friends \citep{DavisM_85a} halos, namely central subhalos, or central galaxies in the case of the hydrodynamical simulations series. This is a conservative choice, as differences between shadow subhalos where one is a central and one is a satellite tend to be larger, due to the strong environment-driven evolution of satellites. Such cases occur when timing differences appear between shadow systems, for example if one, in which the subhalo is still a central, lags behind the other, in which the subhalo is already a satellite. Such cases are quite rare, and the galaxy populations in our simulations are not large enough to sample them well, which is another reason for our choice to exclude them from the main analysis.

\subsubsection{Quantifying the Differences between Shadow Galaxies}
\label{s:analysis_differences}

Once we have a catalog of shadow subhalos between each pair of shadow simulations in a set, we calculate logarithmic differences, namely ratios, in the properties of those shadow subhalos. We focus on the following quantities: total bound stellar mass $M_*$ and dark matter mass $M_{\rm DM}$, the maximum of the circular velocity profile $V_{c,{\rm max}}$ ($\sqrt{GM_{\rm total}/r}$ as a function of radius $r$), the half-mass radius of the stellar distribution $R_{*,1/2}$, the instantaneous SFR based on the gas distribution in the subhalo ${\rm SFR}_0$, and the SFR averaged over a time window of $1\Gyr$, ${\rm SFR}_{1\Gyr}$. All of these quantities are calculated by \SUBFIND{ }during the run, except for ${\rm SFR}_{1\Gyr}$, which we calculate in post-processing based on the formation times of the stellar particles belonging to the subhalo (in Appendix \ref{s:FOFresults} we verify that our results are not significantly affected by the particularities of the \SUBFIND{ }algorithm).

The logarithmic differences of these quantities between shadow subhalos are studied in Section \ref{s:results_individual}. We show that their distributions are well-fit by Gaussians, and quantify the standard deviations of these distributions, namely the typical pairwise differences, as a function of time since the perturbation and of subhalo mass. It is important to realize that the distribution of pairwise differences is wider by $\sqrt{2}$ than the distribution of actual values among many perturbed realizations. This is simply because each realization is drawn from the normal distribution of actual values, and the distribution of pairwise differences is then a distribution of the differences between two identical normal random variables, which is indeed in itself a normal distribution that is $\sqrt{2}$ wider than the original one. In our case, we have a small number of pairwise differences per subhalo, or even just a single one, so we cannot reliably quantify the distribution of actual values. However, we do have a large statistical sample of many galaxies, and therefore many pairwise differences for a population, whose distribution can be robustly quantified and fit with a Gaussian. We therefore present examples of these distributions in and of themselves in \Fig{1d_hists}, as discussed in the next Section. However, it is important to keep in mind that when using the standard deviations of these distributions to quantitatively compare to distributions of {\it values}, rather than of differences, as done in Section \ref{s:results_relations}, the width of the pairwise shadow differences have to be divided by $\sqrt{2}$ for a meaningful comparison, and so this is the way they are presented throughout the paper, with the exception of \Fig{1d_hists}.

In Section \ref{s:results_relations} we go beyond the individual quantities, and study the evolution of differences between shadow galaxies in the context of scaling relations. Specifically, we quantify the extent to which differences in various individual quantities between shadow galaxies move them perpendicular to, versus along, certain scaling relations. To this end, for a given pair of physical quantities, e.g.~stellar mass and halo mass, we perform a piece-wise linear fit in log-space to all the galaxies in all the simulations of a given set. These fits then define the scaling relation between these quantities, as well as the (piece-wise) perpendicular direction to the relation, namely the direction in which the scatter of the relation is minimal. We then calculate the difference between each pair of shadow galaxies in that perpendicular direction. The standard deviation of these pairwise perpendicular differences (divided by $\sqrt{2}$ for reasons discussed in the previous paragraph) is compared to the total scatter (among all galaxies) perpendicular to the scaling relation, in order to assess the contribution of the butterfly effect to the total scaling relation scatter.

\section{Results: Individual Quantities}
\label{s:results_individual}

\subsection{Distributions of Shadow Pairwise Differences}
\label{s:1d_hists}

\begin{figure}
\centering
\includegraphics[width=0.475\textwidth]{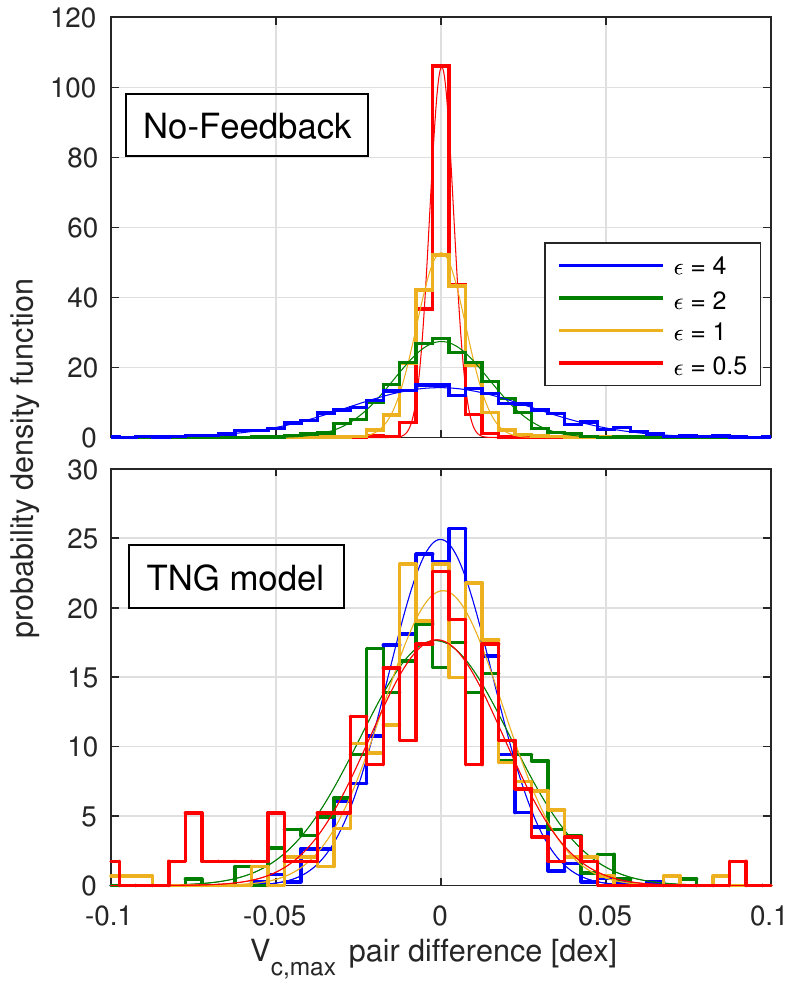}
\caption{Probability density functions of pairwise logarithmic differences between the maximum circular velocities of shadow galaxies with mass of $9.5<\log M_*[h^{-1}\Msun]<10$ at $z=0.5$. These are shown at four resolution levels, increasing from blue to red, for two simulation series, without feedback (top) and with feedback (bottom). The distributions (thick stepwise curves) are fit well by Gaussians (thin curves). Without feedback, the differences between shadow galaxies become smaller as resolution is increased. With the TNG model, however, no clear resolution dependence can be discerned, and the distributions are wider than at high-resolution without feedback.}
\vspace{0.3cm}
\label{f:1d_hists}
\end{figure}

We find that the distributions of pairwise logarithmic differences between the properties of shadow galaxies are well fit by Gaussians whose centers are consistent with zero. This is a general result, which we demonstrate in \Fig{1d_hists} in a particular regime. There, we present the probability density functions of all pairwise logarithmic differences between the values of the maximum circular velocity profile $V_{c,{\rm max}}$ of shadow galaxies, for all central galaxies with stellar mass $9.5<\log M_*[h^{-1}\Msun]<10$ in our No-feedback (top) and TNG model (bottom) simulation series at the last available snapshot, separated by resolution level. In each case, the actual probability density function (thick stepwise curves), which comprises of $\gtrsim100$ of pairwise differences, can be described well by a best-fit Gaussian (thin curves). This shape probably arises due to the central limit theorem, as a large number of individual factors (resolution elements) contribute to the quantity $V_{c,{\rm max}}$. As mentioned, this is a general result that we find holds for other quantities and for other galaxy selections.

The dependence on resolution seen in \Fig{1d_hists} is illuminating. In the No-feedback series, the width of the distribution decreases with increasing resolution: at higher resolution the minute perturbations that are introduced at $z=5$ grow less by $z=0$ than they do at lower resolution. That the result is not converged implies that the magnitude to which these perturbations grow in the lower-resolution cases is not physical, and possibly that their growth is altogether a numerical artifact even in the highest resolution that is available to us, rather than an intrinsic property of the simulated physical system. In particular, as discussed below in Section \ref{s:vs_time_sp}, the results are significantly affected by Poisson noise. In contrast, the results when the TNG model feedback processes are turned on show no meaningful dependence on resolution. At all resolution levels, the standard deviation of the distribution is $\approx0.02\dex$, namely a typical difference of $\approx5\%$ between the $V_{c,{\rm max}}$ values of shadow galaxies. Note that galaxies in the considered mass bin of $9.5<\log M_*[h^{-1}\Msun]<10$ are resolved at the $\epsilon=4$ resolution level with only $\sim30-100$ stellar particles, rendering the invariance of the result between all resolution levels quite striking.

This convergence suggests that the growth of the initial perturbations, on a scale of one part in $10^{14}$, to percent-level differences is inherent to (the numerical realization of) the physical system, namely a system evolving from cosmological initial conditions according to the physical processes included in the TNG model and their particular implementation in this model. In particular, at the $\epsilon=0.5$ resolution level, the distribution of pairwise differences is significantly broader than it is at the same resolution level in the No-feedback case, indicating that the final level of differences is not inherent to the code in general, but is related to the particular physical processes that it implements. Specifically, that the pairwise differences do not keep shrinking with increasing resolution as in the No-feedback case is an indication that the form of feedback implemented in the TNG model increases the sensitivity of the system to small perturbation, or in other words the degree of chaotic-like behavior it manifests.

After establishing that the pairwise differences distributions are Gaussian, throughout the rest of this paper we characterize them with a simple summary statistic: their standard deviation. However, as discussed in Section \ref{s:analysis}, for each individual galaxy, the standard deviation of the pairwise differences between its various shadow versions is a factor of $\sqrt{2}$ larger than the standard deviation of the values themselves. Since here we have only a few pairs per galaxy, we cannot sample the distribution of the values themselves well. However, we have a large number of galaxies, and hence do have a robust estimate of the standard deviation of the distribution of pairwise distances. We hereafter use this robust estimate and divide it by $\sqrt{2}$ in order to obtain a robust estimate of the standard deviation of the values themselves even in the absence of a direct probe into their distribution. As discussed in Section \ref{s:analysis}, the standard deviation of the latter is the more meaningful quantity.

\subsection{Growth of Differences over Time}
\label{s:vs_time}

\subsubsection{Results from No-feedback Simulations}
\label{s:vs_time_sp}

\begin{figure*}
\centering
\includegraphics[width=1.0\textwidth]{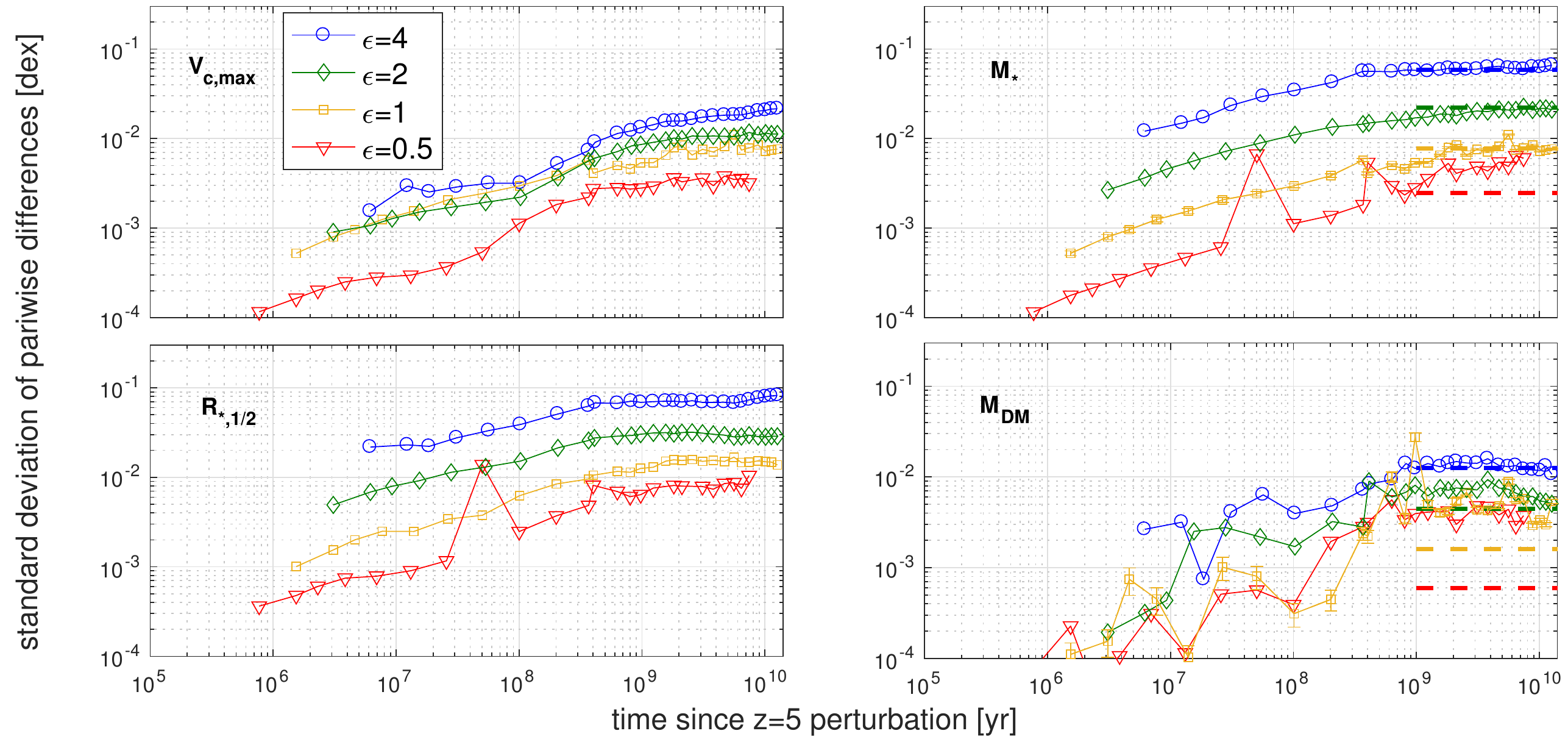}
\caption{The evolution of pairwise differences between shadow galaxies with final mass of $9.5<\log M_*[h^{-1}\Msun]<10$ (in the bottom-right panel: $11.5<\log M_{\rm DM}[h^{-1}\Msun]<12$) in our No-feedback simulation series. Specifically, the standard deviations of the pairwise logarithmic differences distributions (such as those shown in \Fig{1d_hists}), divided by $\sqrt{2}$, are shown as a function of time since $z=5$, when perturbations were applied. Each panel presents these results for a distinct physical quantity: maximum circular velocity, stellar mass, stellar half-mass radius, or halo mass, each based on four resolution levels, which are indicated by color, increasing from blue to red. The results largely show saturation after $\sim1\Gyr$, and a mixture of convergence and non-convergence with resolution. See text for a detailed discussion.}
\vspace{0.3cm}
\label{f:differences_vs_time_SP}
\end{figure*}

\Fig{differences_vs_time_SP} presents the standard deviations of distributions like the ones discussed so far (divided by $\sqrt{2}$, as discussed above) as a function of time, where $t=0$ is defined to be the time the perturbations were introduced, namely in this case $z=5$. These are shown for four physical quantities, one per panel as indicated in the figure, and for four resolution levels via different line styles, as indicated in the legend, all for galaxies from the No-feedback series in a {\it fixed} mass bin of $9.5<\log M_*[h^{-1}\Msun]<10$ (in the bottom-right panel: $11.5<\log M_{\rm DM}[h^{-1}\Msun]<12$). The differences between shadow galaxies have a generic evolution as a function of time for all explored quantities at all resolution levels: an initial growth that can be described reasonably well by a power law $\propto t^{1/2}$, which then plateaus approximately $1\Gyr$ after the perturbation. In other words, after a transition period lasting about $1\Gyr$ after the perturbation, galaxies of $9.5<\log M_*[h^{-1}\Msun]<10$ have a certain (resolution-dependent) degree of random variation between shadow simulations that is independent of cosmic epoch. For most quantities, in accordance with the top panel in \Fig{1d_hists}, the results are not converged, as the differences are smaller at higher resolution, both in the growth phase as well as after reaching a plateau. At the $\epsilon=4$ resolution level, the plateau levels are $\sim0.01-0.1\dex$ for the various quantities, while for the highest, $\epsilon=0.5$ resolution level, they are $\sim0.003-0.01\dex$.

For the two quantities shown on the right column of \Fig{differences_vs_time_SP}, stellar mass $M_*$ (top) and dark matter mass $M_{\rm DM}$ (bottom), there is one source of randomness that is easy to estimate: Poisson noise. Since both stellar and dark matter particles are numerical constructs that discretely sample an underlying smooth field, we can expect random variations on the masses of collections of them, such as subhalos, to scale as $m_p\sqrt{N_p}$, where $m_p$ is the typical particle mass and $N_p$ is the number of particles in a given subhalo. Hence, the relative random scatter in the mass of a subhalo is expected to have a lower limit at $1/\sqrt{N_p}$. These lower limits are shown in the right column of \Fig{differences_vs_time_SP} as horizontal dashed lines. Indeed, this expectation is confirmed, as for the lower resolution levels, the `chaotic' differences between shadow subhalos plateau exactly to the values expected from this Poisson noise estimate. It takes about $1\Gyr$ for the initial perturbations to evolve to that level, since at shorter times after the perturbation the masses of the subhalos still mostly consist of their components that formed prior to the perturbation, and hence is in common to all shadow realizations. In other words, $N_p$ in this context applies to the number of particles added since the perturbation. It is therefore expected that the time to reach the plateau corresponds roughly to the growth timescale of the mass itself, and this is consistent with the observed timescale of $\approx1\Gyr$. Moreover, for a constant mass growth rate ${\rm d}M/{\rm d}t$, which is a reasonable approximation for a relatively short window of $1\Gyr$, $N_p$ is roughly linear with time, and hence the relative error $\sqrt{N_p}/M_*$ (where $M_*$ is a constant by selection) scales roughly as $t^{1/2}$, as indeed observed.

Importantly, the expected Poisson noise diminishes as the square root of the mass resolution, namely by a factor of $\sqrt{8}\approx2.8$ with every step in resolution level. In the case of the stellar mass, the measured `chaotic' differences indeed diminish at that rate for the lowest three resolution levels, indicating that Poisson noise is the dominant factor in those regimes. However, for the $\epsilon=0.5$ level this is no longer the case, as the measured differences are larger than expected from Poisson noise. This indicates that at this high resolution there exists a different origin to the `chaotic' differences that is not just sampling noise. In the case of the dark matter mass, this is even more pronounced, as the differences are larger than expected from Poisson noise at all resolution levels but the lowest one, and in fact the differences appear to be converged between $\epsilon=1$ and $\epsilon=0.5$. This, again, indicates that there is something beyond the simple randomness of the sampling of the mass field that gives rise to mass differences between shadow simulations.

For the quantities shown on the left column of \Fig{differences_vs_time_SP}, maximum circular velocity $V_{\rm c,max}$ (top) and stellar half-mass radius $R_{*,1/2}$ (bottom), it is not clear whether a simple analytic estimate can be devised. It is to be expected that there is an initial growth phase during the time that there still exists a significant component that formed before the perturbation. For reasons that are unknown to us, the differences in $R_{*,1/2}$ grow at a similar rate to those of the masses, roughly $\propto t^{1/2}$, but the growth of the $V_{\rm c,max}$ differences begins slower than that and then accelerates around $10^8\yr$ after the perturbation\footnote{For a possible connection between a rough $\propto t^{1/2}$ divergence of integrated quantities of N-body systems and diffusion, see \citealp{ElZantA_18a}).}. It is also then not entirely expected or straight-forward that the differences between the shadow galaxies in these two quantities reach a plateau around the same time the masses do, $\approx1\Gyr$, suggesting that the differences may be mass-dependent but not time-dependent. Importantly, and curiously, these structural properties that are on the left column show worse convergence than the masses on the right column, suggesting that they might be driven by numerical discreteness that will continue diminishing with increasing resolution.

We conclude the discussion of \Fig{differences_vs_time_SP} with a comment on the statistical uncertainty on these standard deviations. The curves in \Fig{differences_vs_time_SP} are mostly rather smooth, which indicates that the statistical uncertainty is small. Since the distributions from which these standard deviations are measured are to a good approximation Gaussian, the error on the standard deviations can be estimated simplistically by dividing the standard deviation itself by the number of shadow pair differences that constitute the distributions. To avoid visual clutter, we show these simplistic estimates, as error bars, only in the right panels of \Fig{differences_vs_time_SP} and only for the $\epsilon=1$ resolution level, since for this level the uncertainties are the largest as the number of galaxies is the smallest (see Table \ref{t:galaxy_numbers}). This confirms that the statistical uncertainties are similar to the typical point-to-point variations, as expected, and that in most cases these are comparable or smaller than the size of the symbols in \Fig{differences_vs_time_SP}. In Appendix \ref{s:FOFresults} we comment on where this simplistic estimate breaks.

\subsubsection{Results from TNG model Simulations}
\label{s:vs_time_tng}

\begin{figure*}
\centering
\includegraphics[width=1.0\textwidth]{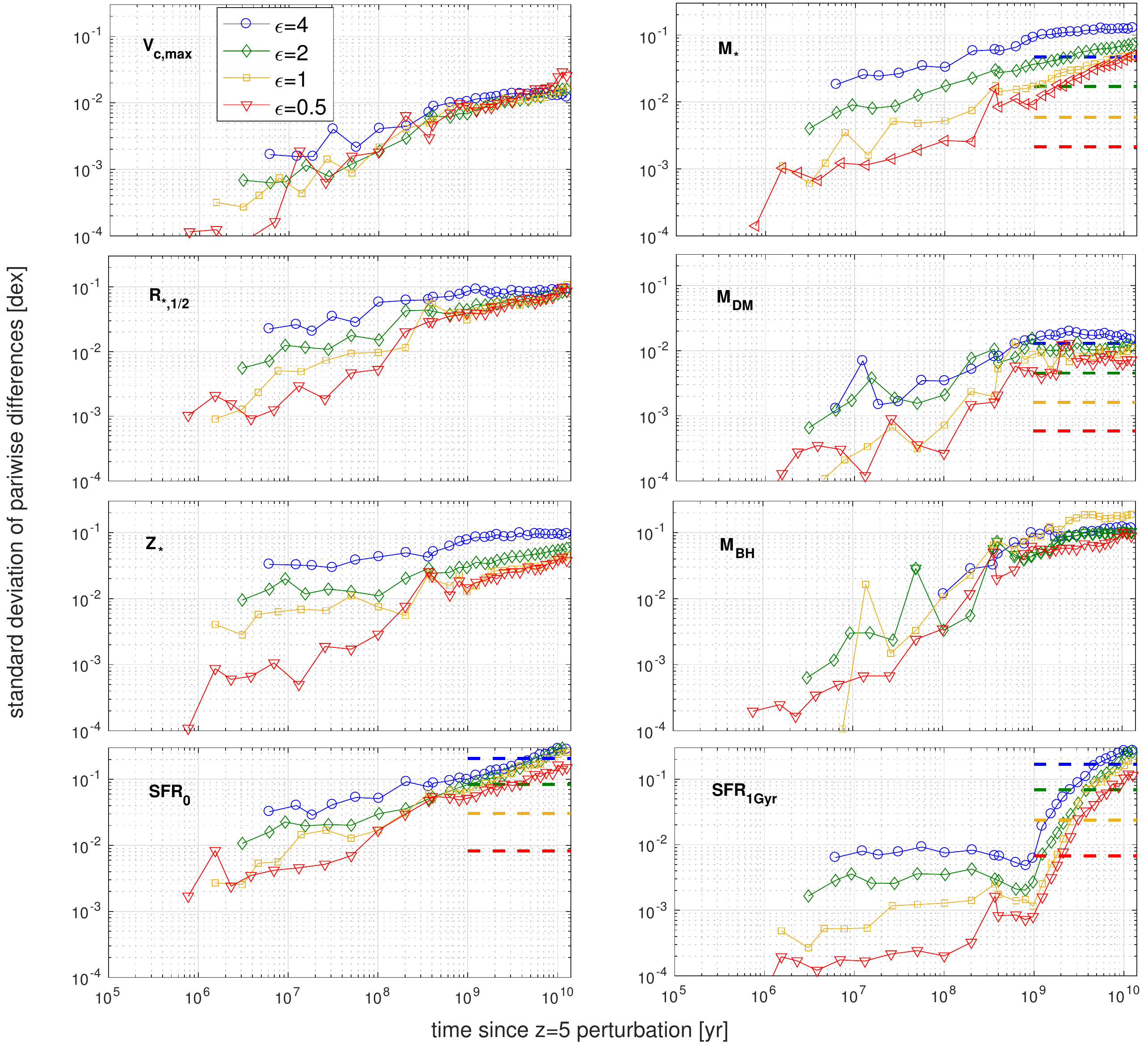}
\caption{The evolution of pairwise differences between shadow galaxies with final mass of $9.5<\log M_*[h^{-1}\Msun]<10$ (in the second from top, right panel: $11.5<\log M_{\rm DM}[h^{-1}\Msun]<12$), similarly to \Fig{differences_vs_time_SP} except here based on our simulation series that uses the TNG model, namely including feedback. In addition to the top four panels that repeat the quantities shown in \Fig{differences_vs_time_SP}, the four bottom panels present additional quantities: stellar metallicity, black hole mass, and SFR measured in two ways. In this case of the TNG model, much clearer convergence is generally seen with increasing resolution (blue to red), compared to the No-feedback case of \Fig{differences_vs_time_SP}. It is also clear that Poisson noise, where it can be straightforwardly estimated (horizontal dashed curves), is very sub-dominant at high resolution.}
\vspace{0.3cm}
\label{f:differences_vs_time_TNG}
\end{figure*}

\Fig{differences_vs_time_TNG} is analogous to \Fig{differences_vs_time_SP} except that it presents the results for the TNG model simulation series, and that it includes four additional panels for additional physical quantities. Several important qualitative differences exist between \Figs{differences_vs_time_SP}{differences_vs_time_TNG}.

First and foremost, the results for the common four physical quantities (top four panels) appear to be well-converged with the TNG model, as opposed to the case without feedback, extending a similar result discussed around \Fig{1d_hists}. In particular, at the highest resolution level, $\epsilon=0.5$, the typical differences among shadow galaxies close to $z=0$ are much larger with the TNG model than without feedback: $\approx0.015\dex$ (or $3.5\%$) versus $\approx0.003\dex$ for $V_{\rm c,max}$ (top-left), $\approx0.05\dex$ (or $12\%$) versus $\approx0.006\dex$ for $M_*$ (top-right), $\approx0.1\dex$ (or $25\%$) versus $\approx0.01\dex$ for $R_{*,1/2}$ (middle-left), and $\approx0.007\dex$ (or $1.5\%$) versus $\approx0.004\dex$ for $M_{\rm DM}$ (middle-right). It appears, then, that the introduction of feedback in the TNG model gives rise to a much stronger amplification of the initial perturbations.

Second, for all the baryonic properties we examine (i.e.~except for $M_{\rm DM}$), the differences appear to be rising with the TNG model at all cosmic times, and in particular still be rising at $z=0$, rather than reaching a plateau as in the No-feedback case. In other words, galaxy mass is no longer the sole determinant of the differences between shadow simulations; instead, galaxies at a fixed mass tend to show a larger effect of the initial perturbations at later epochs.

Third, the evolution of the differences in the stellar and dark matter masses (top-right and second-right panels) is with the TNG model not strongly affected by Poisson noise (the only exception being the $\epsilon=4$ resolution level for $M_{\rm DM}$), but instead continues growing to much higher levels than that, implying that actual physical processes generate these differences rather than effects of discrete sampling. Interestingly, the growth keeps its approximate power-law dependence on time even after crossing the maximal Poisson noise level, namely that which corresponds to the (full, rather than accreted/formed after the perturbation) particle number in the selected mass bin. This is curious as the explanation we suggested above for this dependence applied only to the early regime, before reaching that level.

The bottom four panels in \Fig{differences_vs_time_TNG} present four additional quantities that were not included in \Fig{differences_vs_time_SP} for the No-feedback model. In the third row on the left are the logarithmic differences between shadow galaxies in stellar metallicities. The results are systematically converging and appear very well converged between the two highest resolution levels after $\sim1\Gyr$, at a level of $\approx0.04\dex$ (or $10\%$) at $z=0$. In the third row on the right are the differences in black hole masses, which hover around $\approx0.1\dex$ (or $25\%$) at $\gtrsim1\Gyr$ for the various resolution levels, which however do not show a monotonic behavior, as discussed below.

The two quantities examined in the bottom row of \Fig{differences_vs_time_TNG} are measurements of the SFR, but on different timescales. In the bottom-left, it is the instantaneous SFR as measured from the gas cells, which is determined by their density based on the \citet{SpringelV_03a} model, ${\rm SFR}_0$. In the bottom-right, it is the SFR averaged over the past $1\Gyr$, as measured from the number of stellar particles that actually formed during this time window, ${\rm SFR}_{1\Gyr}$. For both quantities, an estimate of Poisson errors can be made based on the number of resolution elements that contribute to the calculation. For ${\rm SFR}_0$ this is somewhat less accurate, as the instantaneous SFRs of individual cells can vary greatly, than for ${\rm SFR}_{1\Gyr}$, which is based on the almost-constant masses of stellar particles. Nevertheless, both quantities show a similar picture indicating that Poisson noise\footnote{Unlike mass, which is constant by selection, the SFRs change over cosmic time, and hence the Poisson noise level is not constant. The dashed horizontal lines in the bottom two panels of \Fig{differences_vs_time_TNG} are calculated based on the $z=0$ SFRs, which are at their nadir at that time, resulting in larger Poisson noise levels than at any other cosmic epoch.} does not dominate, except perhaps at the lowest resolution level\footnote{Note that the feature at $1\Gyr$ that appears for ${\rm SFR}_{1\Gyr}$ is there essentially by construction, as at all times shorter than $1\Gyr$ past the perturbation, the measurement of ${\rm SFR}_{1\Gyr}$ is based partially on stellar particles that were formed prior to the perturbation, namely ones that are by construction in common between all the shadow simulation in a set. Only after longer times can and do the differences grow substantially to (and even beyond) the indicated Poisson noise level, which is calculated assuming that {\it all} the particles are independent draws from some underlying smooth field.}. The effect of the perturbations is clearly still rising for the SFRs as a function of cosmic time even at $z=0$, for galaxies in this fixed mass bin. Perhaps surprisingly, the differences between shadow simulation in ${\rm SFR}_{1\Gyr}$ are quite close to those in the instantaneous ${\rm SFR}_0$, both being $\approx0.2\dex$ at $z=0$. This indicates that the star formation histories of shadow galaxies diverge from one another in a significant way not only on short timescales, but rather even when averaged over time windows much longer than, e.g., a galactic dynamical time. It is also worth pointing out, for context, that this level of differences between shadow galaxies is comparable to the overall scatter of SFRs between galaxies in this stellar mass bin, a point discussed in more detail in Section \ref{s:results_relations}.

The results in the bottom row of \Fig{differences_vs_time_TNG} show a curious behavior with respect to dependence on resolution. They appear essentially converged (at $\gtrsim1\Gyr$) between the two intermediate resolution levels of $\epsilon=2$ and $\epsilon=1$, but then diverge toward smaller values for the highest level, $\epsilon=0.5$. We interpret this as evidence that star formation itself proceeds in a different way in the $\epsilon=0.5$ set, affecting the process of perturbation amplification. This is in accordance with the findings of \citet{SparreM_14a} that a new, more bursty mode of star-formation appears at resolution levels beyond that of Illustris. In other words, not only the process we study here, namely the perturbation amplification, is affected by changing resolution, but also the results of the simulation itself and thereby also the dominance and effect of various physical processes that occur within the simulation and which drive the perturbation amplification. It is hard to separate the direct effect of numerical resolution on the perturbation amplification from its indirect effect through changes to the simulation results themselves and to the relevant physical processes. In fact, this indirect effect of resolution is not guaranteed to act in the direction of decreasing the amplification. Indeed, a non-monotonic dependence on resolution appears for the case of black hole masses (third from top panel on the right), and a tentative hint for an opposite effect can be seen in the top-left panel for $V_{c,{\rm max}}$, where at late times the growth of the differences is faster, and their amplitude is larger, in the $\epsilon=0.5$ case than in the other resolution levels, which in themselves appear quite converged. A careful examination of \Fig{1d_hists} reveals that, at least in the last snapshot, this is driven by the larger number of outliers in the $\epsilon=0.5$ case, which one may speculate to, as well, be driven by the more bursty mode of star formation at this resolution level. While this particular case is not conclusive due to small number statistics, this general point is discussed further in Section \ref{s:vs_mass}.

\subsubsection{Results of Simulations Without Random Numbers}
\label{s:no_random}

\begin{figure*}
\centering
\includegraphics[width=1.0\textwidth]{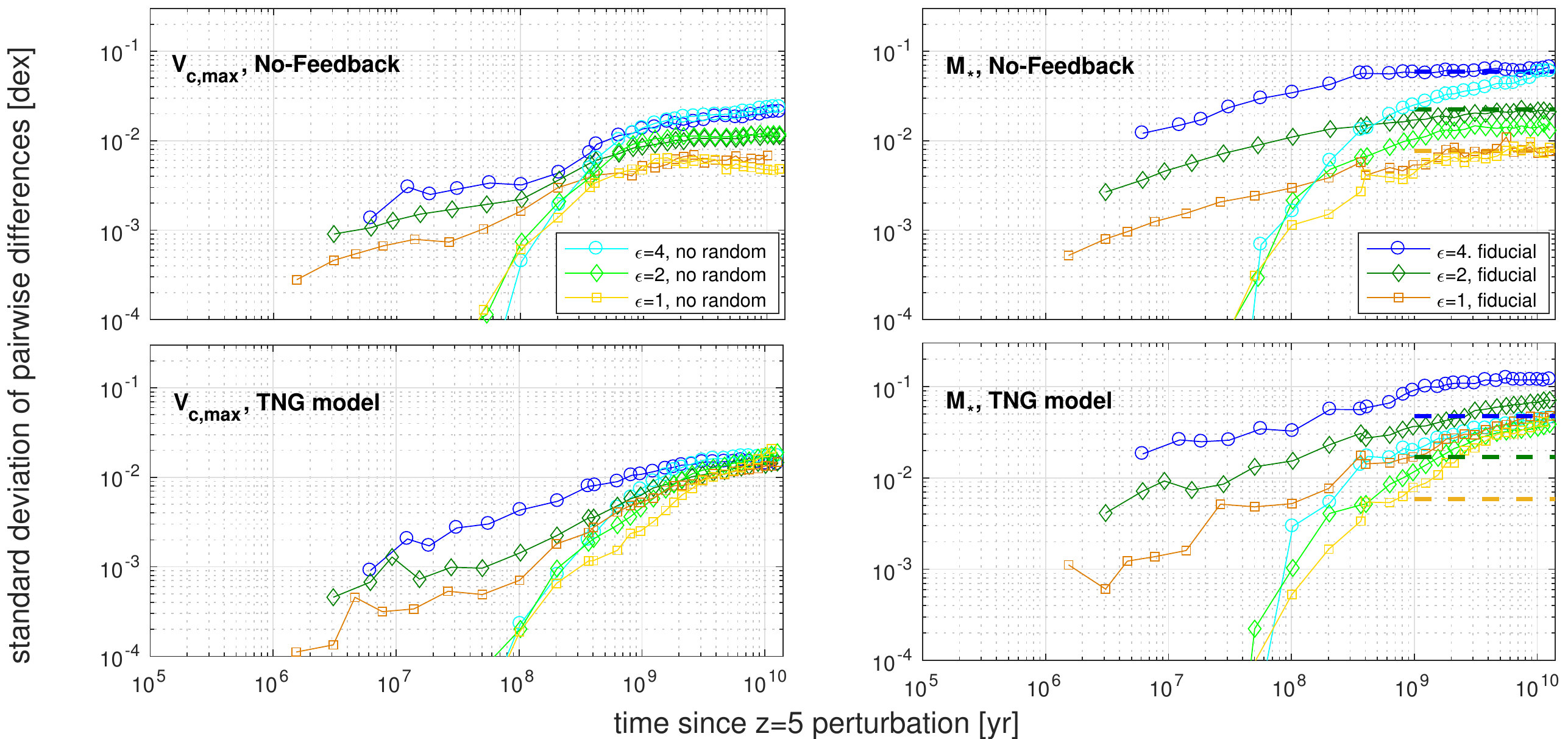}
\caption{The evolution of pairwise differences between shadow galaxies with final mass of $9.5<\log M_*[h^{-1}\Msun]<10$, similarly to \Fig{differences_vs_time_SP} (No-feedback, top) and \Fig{differences_vs_time_TNG} (TNG model, bottom), but here comparing the fiducial models (dark colors) to modified subgrid models that completely avoid the usage of random numbers (light colors). Two quantities are presented: maximum circular velocity (left) and stellar mass (right). In almost all cases (see discussion in Section \ref{s:no_random}), the differences evolve more gradually at early times ($t\lesssim1\Gyr$) in the simulations without random numbers, but eventually converge to very similar values as in the fiducial simulations. This demonstrates that the butterfly effect in cosmological simulations is not driven by the usage of random numbers in the subgrid models.}
\vspace{0.3cm}
\label{f:no_rand_sims}
\end{figure*}

In \Fig{no_rand_sims} we demonstrate that the results presented thus far are largely unchanged when the usage of random numbers in the simulations is turned off. For both the No-feedback (top) and the TNG models (bottom), the growth of differences is compared for two quantities, $V_{\rm c,max}$ (left) and $M_*$ (right), between the fiducial simulations (dark colors) and the simulations run with modified subgrid models that do not use random numbers (light colors). Two general trends visible in this comparison stand out.

First, at early times the evolution of the differences between the two types of simulations is markedly different. Specifically, in the fiducial simulations the differences appear at a level of $\sim10^{-3}$ already after a few million years, in the first snapshots that are available. Thereafter, the evolution is rather gradual, with a power-law behavior as discussed in the previous sub-sections. In contrast, it takes $\gtrsim100\Myr$ for the simulations without random numbers to reach this level: their evolution in the first few million years is much slower, and thereafter is much faster. In Appendix \ref{s:verification} we discuss in much more detail the very early evolution, and how it can be dominated by the usage of random numbers. To summarize the conclusions from Appendix \ref{s:verification}, the differences in random number sequences that develop between pairs of shadow fiducial simulations result in a `discontinuous' evolution of the pairwise differences. This is avoided when random numbers are not used, resulting in an exponential growth of the initial differences with (Lyapunov) timescales on the order of the dynamical time of galaxies at the perturbation redshift of $z=5$. This exponential growth is more gradual than the `discontinuous' initial growth in the fiducial simulations but is faster thereafter.

Second, after enough dynamical times, the evolution in the simulations without random numbers catches up and the pairwise differences converge to values that are essentially indistinguishable from those in the fiducial simulations. This indicates that the late-time ($\gtrsim1\Gyr$) evolution of the pairwise differences is roughly independent of how they are `seeded' at earlier times, namely either by a power-law growth of early `discontinuous' differences brought about by random number differences, or by an exponential growth of the perturbations introduced initially. One regime where this is not the case is the stellar mass in lower-resolution simulations with the TNG model (bottom right panel), where the plateau level of pairwise differences is at larger values in the fiducial set than in the set without random numbers. It appears that in these lower-resolution cases the use of random numbers increases the pairwise differences. These are reduced as the resolution increases, such that the fiducial TNG model is not yet converged between the resolution levels shown in \Fig{no_rand_sims}. In contrast, the simulations without random numbers show converged results at late times at a level that is in fact very similar to the converged results of the fiducial simulations (seen also in the top right panel of \Fig{differences_vs_time_TNG}).

\subsubsection{Comparing Shadow Differences to Overall Scatter}
\label{s:cos_alpha}

\begin{figure}
\centering
\includegraphics[width=0.475\textwidth]{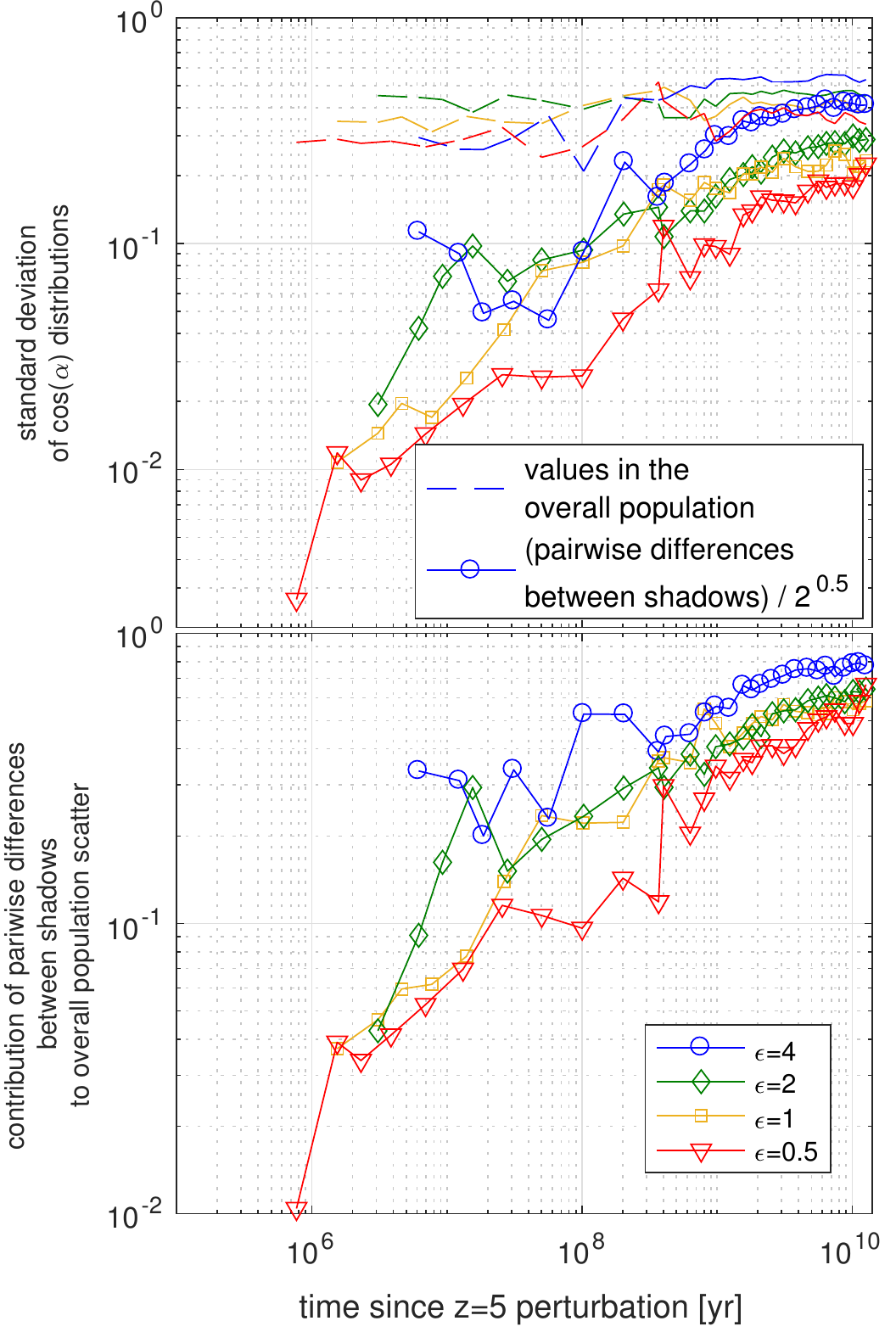}
\caption{{\it Top:} a comparison of the standard deviations of the distributions of pairwise ${\rm cos}(\alpha)$ differences between shadow galaxies (divided by $\sqrt{2}$; solid curves with symbols) to the standard deviations of the ${\rm cos}(\alpha)$ distributions of the overall galaxy population (dashed curves), where $\alpha$ is the angle between the angular momentum vectors of the stellar and total mass contents of the \SUBFIND{ }subhalos hosting the galaxies. The comparison is made as a function of time since the perturbation is applied at $z=5$, and includes all central galaxies with stellar mass above $10^{8.5}h^{-1}\Msun$. {\it Bottom:} the ratio between the two quantities shown in the top panel.}
\vspace{0.3cm}
\label{f:differences_vs_time_angle}
\end{figure}

We close this subsection with a study of one additional quantity, the angle between the angular momentum vector of the stellar component of subhalos and that of their total mass content (including the dark matter and gas), which we denote $\alpha$. Pairwise differences of ${\rm cos}(\alpha)$ between shadow galaxies are presented in \Fig{differences_vs_time_angle} (for this quantity, we find no mass dependence, hence this figure is based on all galaxies with $M_*>10^{8.5}h^{-1}\Msun$). In the top panel, the solid curves are analogous to those in \Figs{differences_vs_time_SP}{differences_vs_time_TNG}, and they present a similar picture of an initial power law-like growth and a plateau reached at $t\gtrsim1\Gyr$, which is resolution-dependent, but possibly close to converged in the highest resolution level.

In addition, the top panel in \Fig{differences_vs_time_angle} shows (dashed curves) the standard deviations of the distributions of ${\rm cos}(\alpha)$ values of different galaxies, rather than ${\rm cos}(\alpha)$ differences between shadow galaxies (solid curves). These ${\rm cos}(\alpha)$ values are of all galaxies with $M_*>10^{8.5}h^{-1}\Msun$ in all of the simulations of any given resolution level, combined, but practically indistinguishable standard deviations are obtained when only a single (arbitrary) simulation is used (for any given resolution level). To emphasize, this quantity, the standard deviation of the distribution of the values of a certain property, is the quantity that is regularly being referred to as the overall scatter in this property, in this case ${\rm cos}(\alpha)$. It is seen to be rather constant as a function of time and for the most part between the four resolution levels, at $\approx0.3-0.5$. At low resolution, however, the overall ${\rm cos}(\alpha)$ scatter is larger than at higher resolutions, and is not much larger than the typical difference between shadow galaxies (solid curves). This suggests that it is the butterfly effect itself that affects, namely enhances, the overall ${\rm cos}(\alpha)$ scatter at the $\epsilon=4$ level. When the former drops, at higher resolution, so does the latter.

Shown in the bottom panel of \Fig{differences_vs_time_angle} is the ratio between the solid and dashed curves of the top panel, which can be interpreted as the fractional contribution of the butterfly effect to the total scatter in this quantity. Since we compare standard deviations of distributions, and plausibly other contributions of scatter would be independent and hence add quadratically, it is the square of the ratio shown in the bottom panel that is the more meaningful quantity, namely the contribution of the butterfly effect to the {\it variance} of ${\rm cos}(\alpha)$ among the overall galaxy population. In the highest-resolution case it appears possibly converged at $0.5^2$, which implies that about $25\%$ of the variance among galaxies in the misalignment between these two vectors cannot be derived from deterministic macroscopic arguments -- which is to say, that portion of the variance cannot be predicted or explained. In Section \ref{s:results_relations} we will make similar comparisons, but for the scatter of a scaling relation between two quantities instead of for the scatter in an individual quantity.

\subsection{Differences versus Mass}
\label{s:vs_mass}

\begin{figure*}
\centering
\includegraphics[width=1.0\textwidth]{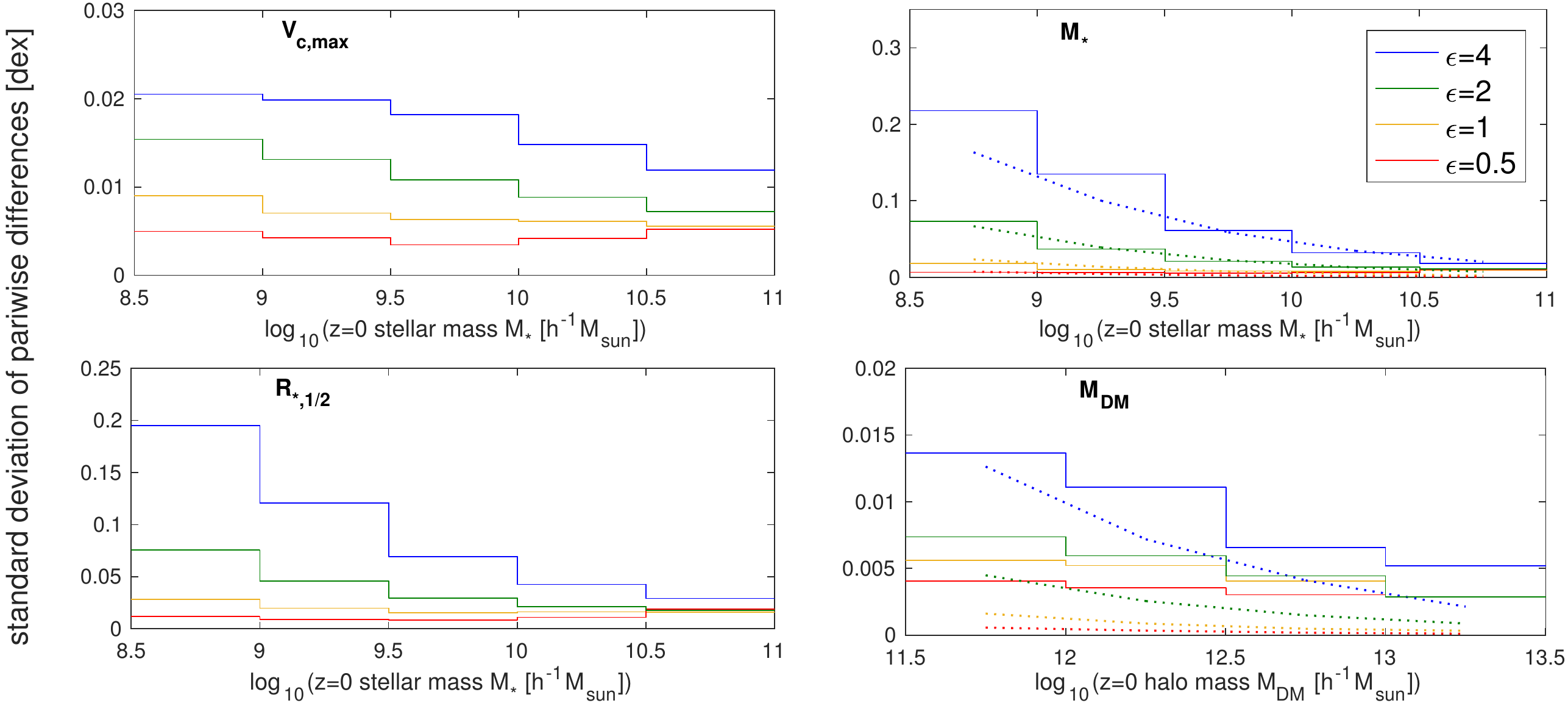}
\caption{Pairwise differences (specifically: the standard deviations of the distributions thereof, divided by $\sqrt{2}$) between shadow galaxies in our No-feedback simulation series, as a function of final mass, averaged over the six snapshots in the redshift range $0.5\leq z\leq 1.5$. Each panel presents these results for a distinct physical quantity: maximum circular velocity, stellar mass, stellar half-mass radius, or halo mass, each based on four resolution levels, which are indicated by color, increasing from blue to red.}
\vspace{0.3cm}
\label{f:vs_mass_SP}
\end{figure*}

In Section \ref{s:vs_time} we have shown how differences between shadow simulations grow as a function of time for a fixed selected mass bin. Here, we present a complementary view, of the late-time differences between shadow subhalos of various mass bins (all perturbed with respect to one another at $z=5$). These results are shown in \Fig{vs_mass_SP} for the No-feedback series, in an analogous organization to that of \Fig{differences_vs_time_SP}, with a different physical quantity in each panel and a different color for each resolution level. The middle of the five stellar mass bins, $9.5<\log M_*[h^{-1}\Msun]<10$, is the one for which results were discussed in Section \ref{s:vs_time}, and so is the first of the four dark matter mass bins in the lower-left panel, $11.5<\log M_{\rm DM}[h^{-1}\Msun]<12$. In order to increase the statistical significance, the results are shown for the average of six snapshots corresponding to the redshifts of the last six snapshots available for the highest resolution level, which in the case of the No-feedback series corresponds to $0.5\leq z\leq 1.5$.

The top-right panel in \Fig{vs_mass_SP} shows that the logarithmic differences between the stellar masses of shadow galaxies in the No-feedback series is strongly mass-dependent, and specifically smaller for more massive galaxies. This is easy to understand, as the close match is apparent between the actual data (solid steps) and the estimates based on Poisson noise (dotted curves). Hence, the conclusion from the top-right panel of \Fig{differences_vs_time_SP} regarding the Poisson noise origin of the differences holds generally for all mass bins. The exception to this conclusion, which is also in alignment with \Fig{differences_vs_time_SP}, is the highest resolution level, and in particular so, for higher mass bins. In particular, for galaxies with $\log M_*[h^{-1}\Msun]>10$, the stellar mass differences between the shadow $\epsilon=0.5$ simulations are several times larger than expected based purely on sampling noise given the number of particles comprising these galaxies. Still, the standard deviations between the stellar masses in shadow galaxies at this high resolution level is rather small, $\approx0.01\dex$, across the $8.5<\log M_*[h^{-1}\Msun]<11$ mass range.

The result is quite different for the dark matter mass of these subhalos, as shown in the lower-right panel of \Fig{vs_mass_SP}. It is not clear if the results show convergence toward a value larger than zero, however they are definitely larger than expected purely due to Poisson noise, at all resolution levels and all masses, except at the combination of lowest mass and lowest resolution. Nevertheless, the lower resolution levels, and in particular $\epsilon=4$, are clearly affected by the Poisson noise to a certain degree. At the end of the day, the magnitude of the result at the highest resolution level may be considered small: it is $\approx0.005\dex$ across the full mass range explored.

The results on the left column of \Fig{vs_mass_SP} do suggest convergence toward mass-independent values of $\approx0.005\dex$ for the maximum circular velocity (top) and $\approx0.01\dex$ for the stellar half-mass radius (bottom). We do not have an analytical estimate analogous to the one we have for the mass-based quantities shown on the right, but it is nevertheless clear that at lower simulation resolution levels, lower-mass galaxies are more strongly affected by the butterfly effect, and that this mass dependence becomes weaker at higher resolutions. This suggests that with regards to these quantities too there is a role for discreteness or sampling effects. These effects however appear to be largely mitigated at the $\epsilon=0.5$ resolution level, where a mass-independent `floor' is reached.

\begin{figure*}
\centering
\includegraphics[width=1.0\textwidth]{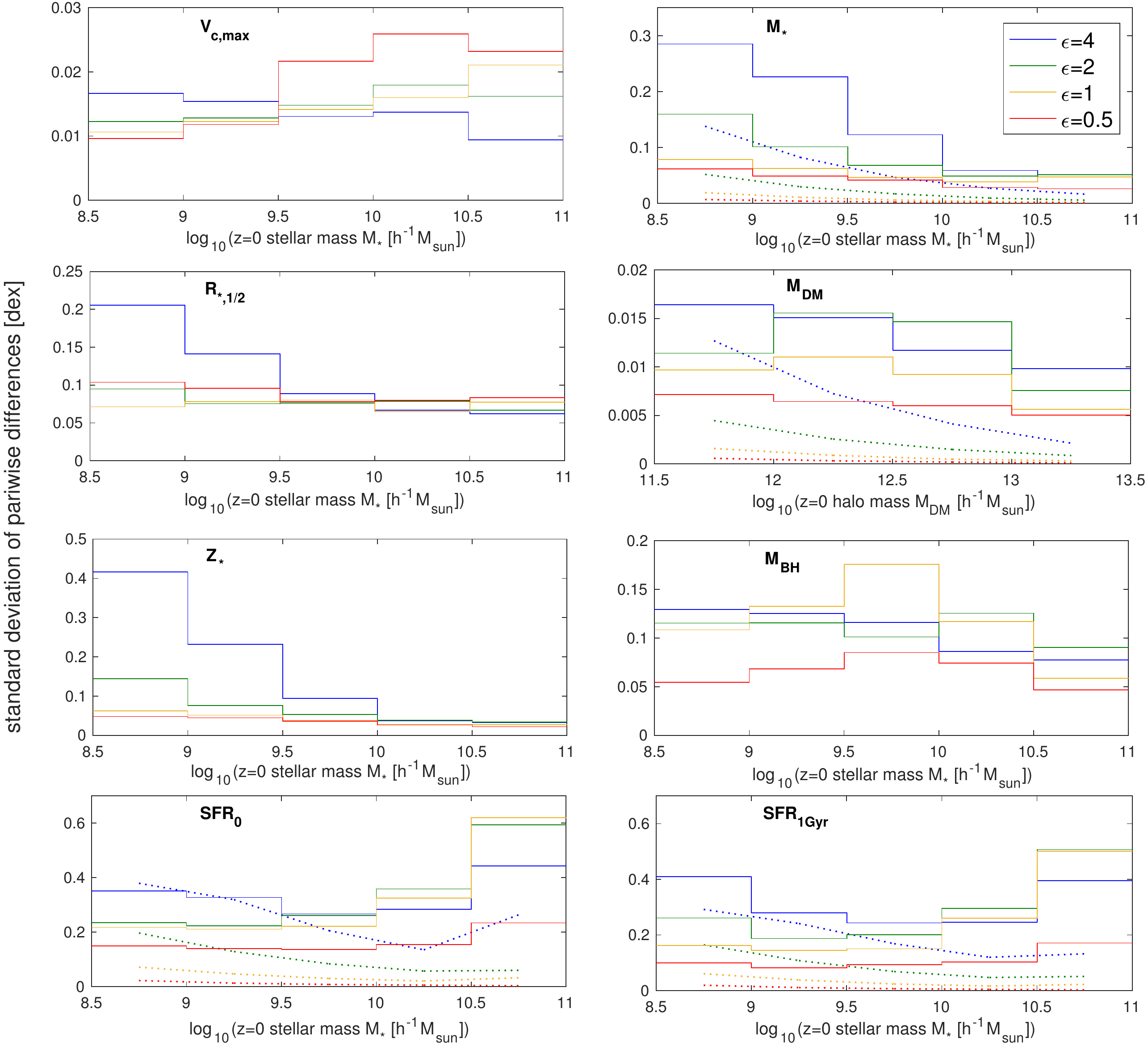}
\caption{Pairwise difference between shadow galaxies as a function of final mass, similarly to \Fig{vs_mass_SP}, but for the simulation series based on the TNG model, namely including feedback, and with the addition of four measurements corresponding to the bottom half of \Fig{differences_vs_time_TNG}. Here the six snapshots that are included cover $0\leq z\leq 0.65$.}
\vspace{0.3cm}
\label{f:vs_mass_TNG}
\end{figure*}

In \Fig{vs_mass_TNG} we present a similar study, but for the TNG-model simulation series, where again the structure is analogous to that of the time-dependent \Fig{differences_vs_time_TNG}. The phenomenology seen in \Fig{vs_mass_TNG} is quite rich, and we here discuss the aspects we find most significant and illuminating.

\begin{itemize}
\item As seen in the top-left panel of \Fig{differences_vs_time_TNG} for the middle mass bin shown here, the dependence of the $V_{c,{\rm max}}$ differences on resolution is not monotonic, and while the results for three resolution levels are very close to each other, those for the highest one are markedly different. The examination here of additional mass bins reveals a more general picture: in the low mass bins, higher resolution results in lower differences, while in the high mass bins, higher resolution results in {\it larger} differences. This highlights an argument made in the discussion of \Fig{differences_vs_time_TNG}, namely that changes with resolution may arise due to the appearance of new physical processes or phenomena at higher resolution levels, for example in the mode of star formation, or the galactic dynamics. This highlights further the idea that our quantitative results are idiosyncratic to the particular physical model that is employed, in the broadest sense that involves also the numerical resolution, and cannot immediately be generalized to other numerical or physical setups, or to the real universe. A similar discussion is relevant for the behavior of black hole mass differences (third row, right column).
\item The results for stellar mass (top-right) are similar for all mass bins we consider except for the highest one. The stellar mass differences are significantly larger than those expected purely from Poisson noise, and instead decrease with increasing resolution in a way that appears to converge toward a finite value that is only mildly mass-dependent. Specifically, in all mass bins and resolution levels, when galaxies are represented by more than roughly $100$ stellar particles, the differences between shadow simulations become almost independent of the number of particles (even up to $\sim10^5$ particles), and are typically $\sim0.03-0.05\dex$. An exception to the appearance of convergence is the highest mass bin, which shows a rather sharp decrease between the three low resolution levels and the highest one. This is possibly related to the decreased scatter in the high-mass, high-resolution case shown in the bottom two panels, discussed next.
\item At the highest resolution level, the differences between shadow simulations show a nearly mass-independent value, $\approx0.12\dex$ for ${\rm SFR}_0$ (bottom-left) and $\approx0.08$ for ${\rm SFR}_{1\Gyr}$ (bottom-right). At first glance (perhaps surprisingly), in the lower mass bins this appears to be a value toward which the lower resolution levels are converging, while in the higher mass bins the results are non-monotonic with resolution, and in particular show a large decrease between the three lower resolution levels and the highest one. We hypothesize that this has to do with the onset of quenching in the high mass bins, and its sensitivity to resolution. In particular, if it is the case that the butterfly effect can determine whether a galaxy is quenched or not, large shadow pairwise differences are to be expected. Since the quenched fraction is high in high mass bins (e.g.~\citealp{NelsonD_17a}), it should not be surprising that the differences are indeed seen to increase with mass. This is not the case, however, for the highest resolution level, where the results are more in line with the lower mass bins, potentially indicating weaker quenching at this high resolution. To test this hypothesis, we calculate the mean and width of the SFR distributions of all galaxies (not of differences between shadow galaxies) in the two highest mass bins at the $\epsilon=0.5$ resolution level. We find $1.9\Msunyr$ and $3.8\Msunyr$ for the means and $0.23\dex$ and $0.4\dex$ for the scatters, for the two bins respectively. For the $\epsilon=1$ resolution level, in contrast, strong quenching exists in these high mass bins, where the means are $1.3\Msunyr$ and $0.9\Msunyr$ and standard deviations $0.55\dex$ and $1.1\dex$, respectively. This indeed serves as evidence in support of our hypothesis.
\item The results for the half-mass radius (second row, left) are remarkably insensitive to resolution variations and show little mass dependence, with a standard deviation of $\approx0.07\dex$ across this parameter space. The exceptions are low-mass bins at the lowest resolution, which contain only a few dozen stellar particles and hence show larger differences. Those however quickly reach their converged values already at the $\epsilon=2$ resolution level. This is to say, all galaxies at all resolution levels that are resolved by more than $\approx20$ particles show a roughly converged result. The results for the stellar metallicities (third row, left) are the most well-behaved with resolution, showing both a monotonic and converging trend of decreasing differences as the resolution increases, and in particular results that are very similar between the two highest resolution levels.
\end{itemize}

\section{Results: Scaling Relations}
\label{s:results_relations}
\begin{figure}
\centering
\subfigure[Joint distribution of maximum circular velocity and stellar mass]{
          \label{f:2d_hists_TFR}
          \includegraphics[width=0.48\textwidth]{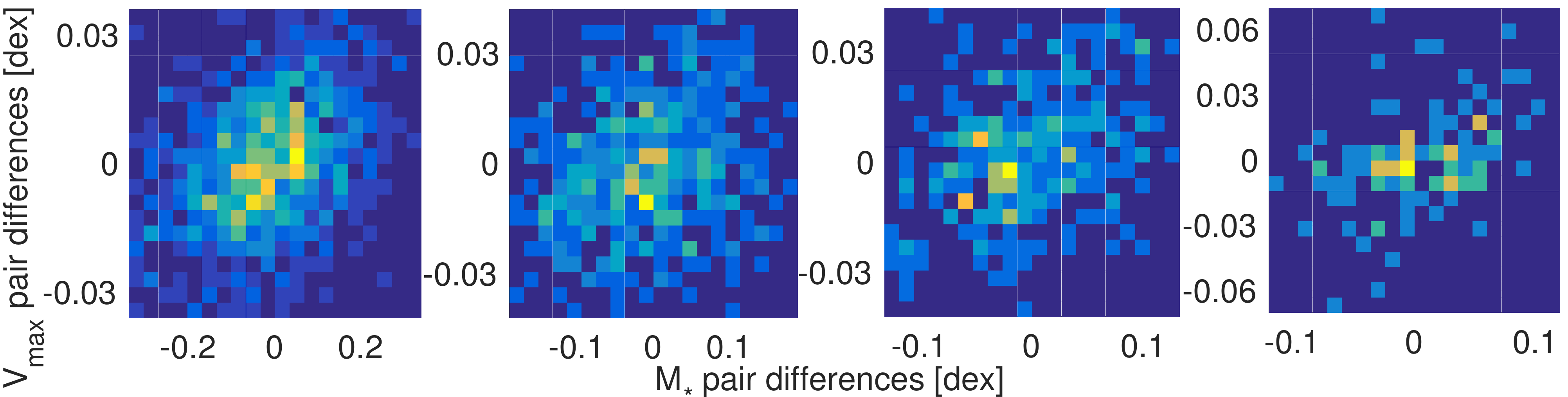}}
\subfigure[Joint distribution of SFR and stellar mass]{
          \label{f:2d_hists_SFRM}
          \includegraphics[width=0.48\textwidth]{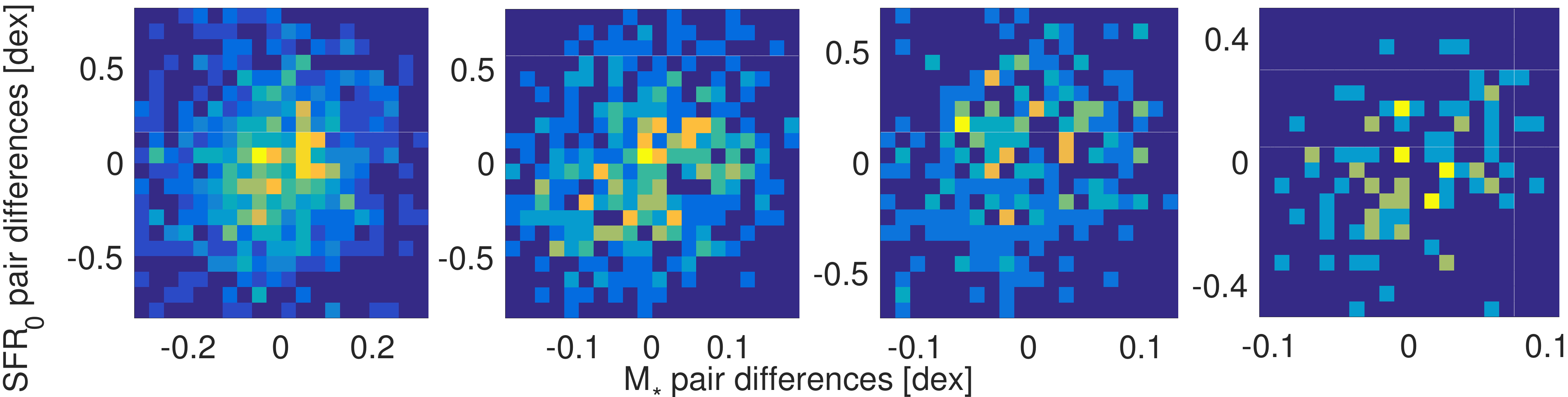}}
\subfigure[Joint distribution of stellar half-mass radius and stellar mass]{
          \label{f:2d_hists_RM}
          \includegraphics[width=0.48\textwidth]{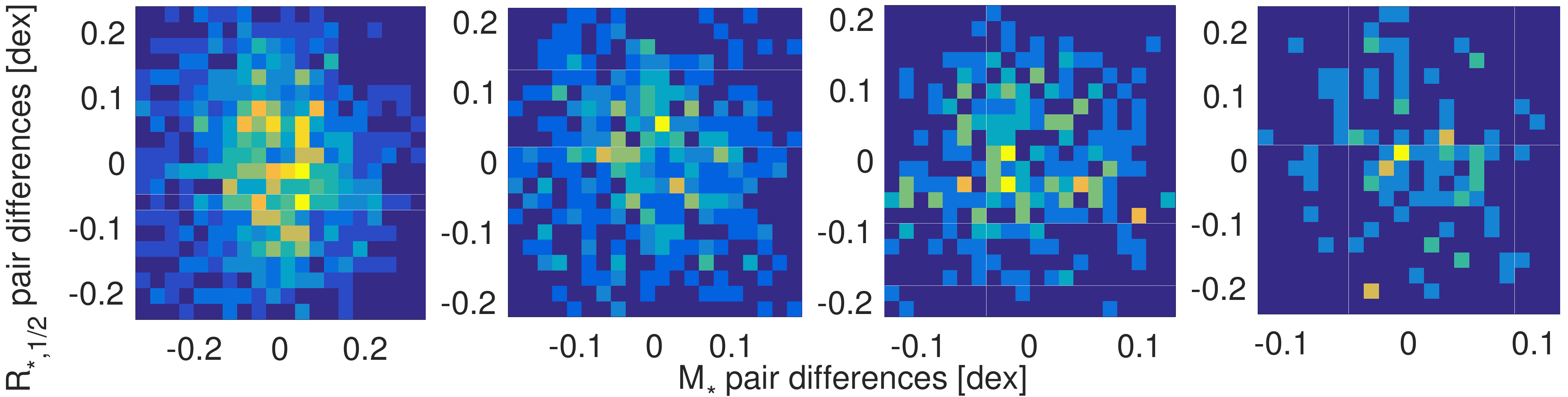}}
\subfigure[Joint distribution of dark matter and stellar masses]{
          \label{f:2d_hists_MsMh}
          \includegraphics[width=0.48\textwidth]{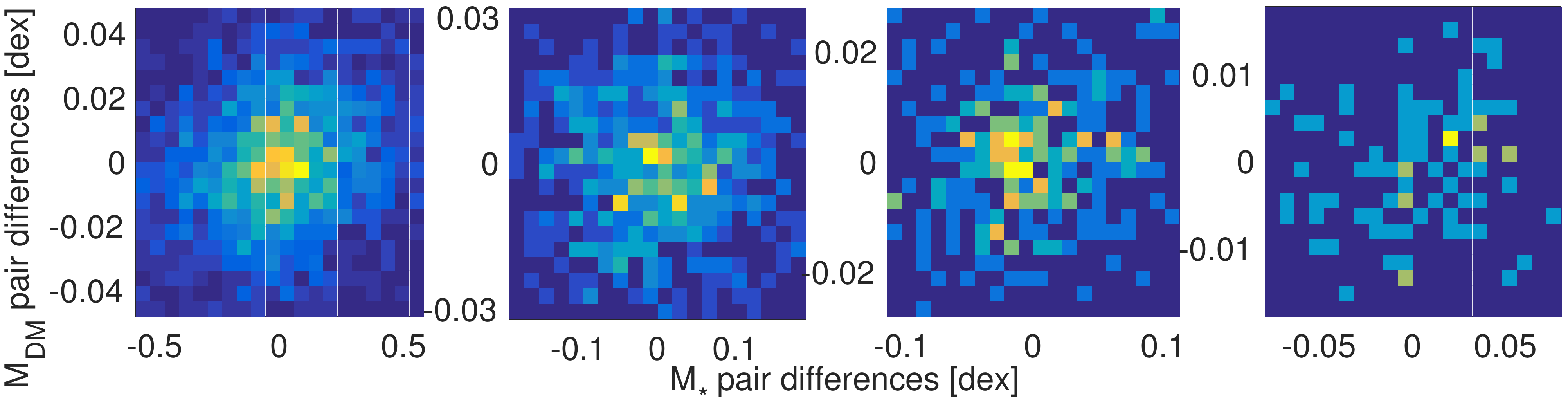}}          
\caption{Joint distributions of shadow pairwise differences in various combinations of two physical quantities. These results are based on $z=0.2$ galaxies with $9.5<\log M_*[h^{-1}\Msun]<10$ (top three rows) or $11.5<\log M_{\rm DM}[h^{-1}\Msun]<12$ (bottom row), at four resolution levels increasing from left to right, all using our simulation series based on the TNG model. The pairwise differences between $V_{c,{\rm max}}$ and $M_*$ (first and second rows, respectively) appear to be somewhat positively correlated albeit with large scatter, while those between $R_{*,1/2}$ and $M_*$ (third row) tend to be very slightly anti-correlated, and the $M_{\rm DM}$-$M_*$ differences show no discernable correlation at all. The total width of each panel in each axis equals to four standard deviations of the one-dimensional distribution of the quantity shown on that axis. Note that the distributions are better sampled at lower resolutions because of the larger number of available shadow galaxy pairs, a trend driven by computing power (see Table \ref{t:simulations}).}
\vspace{0.3cm}
\label{f:2d_hists}
\end{figure}

While so far we have quantified and discussed the differences that develop between shadow simulations one physical quantity at a time, we now turn to study relations between the differences in pairs of quantities, and the implications of those for our general understanding of `galaxy scaling relations', namely correlations between several quantities within a population of galaxies. We begin by presenting an extension into two dimensions of \Fig{1d_hists}, which presented examples of one-dimensional distributions of pairwise logarithmic differences between shadow galaxies. In \Fig{2d_hists} we show several examples of how these differences in one quantity are related to those in another, using heat maps that represent the two-dimensional distributions of differences in several such pairs of quantities. These are all based on the $z=0.2$ snapshot in the TNG-model series of simulations that have been perturbed at $z=5$. Each row shows a different combination of two quantities, with one panel per resolution level, increasing from left ($\epsilon=4$) to right ($\epsilon=0.5$).

The first row of \Fig{2d_hists} demonstrates that the differences between shadow galaxies in stellar mass and maximum circular velocity are positively correlated with substantial scatter. The situation is similar between stellar mass and SFR (second row). On the other hand, there appears to be a very mild anti-correlation between stellar mass and half-mass radius differences (third row), and no significant correlation between stellar mass and dark matter mass (bottom row). These (non/)correlations appear to be stable with resolution variation even as the magnitudes of the differences themselves vary significantly in some cases (in particular, $M_*$ and $M_{\rm DM}$). We do not aim here to explain these results, but we discuss their implications.

If the differences between shadow galaxies in a pair of quantities relate to each other in a similar way to the mean relation between those quantities for a large galaxy population, then we can say that these two shadow galaxies are displaced with respect to one another `along' the overall `scaling relation' between those quantities.
This holds also for the case of an anti-correlation that goes exactly in the opposite direction.
If, however, the differences relate to each other in a different way, then the line connecting the two shadow galaxies is not parallel to the overall scaling relation, and there is a component that is perpendicular to it and parallel to its scatter.

If, for example, the differences between two quantities are uncorrelated at the galaxy population level, then the displacements between pairs of individual shadow galaxies would tend to have some non-zero component perpendicular to the scaling relation between those two quantities. Some pairs would be displaced perpendicular to the relation, some parallel to it, and most in some intermediate direction. \Fig{2d_hists} clearly indicates that that is the case for the pairs of quantities shown in the third and fourth rows, where there is no significant correlation between the differences, even though the quantities themselves certainly are correlated. However, there is significant scatter even in the case of the first and second rows, which do show some overall positive correlation that is indeed similar to the overall scaling relation between the two quantities. Hence, also in the case of the $M_*$-$V_{c,{\rm max}}$ plane, individual pairs of shadow galaxies are expected to show significant displacements in all directions.

\begin{figure*}
\centering
\includegraphics[width=1.0\textwidth]{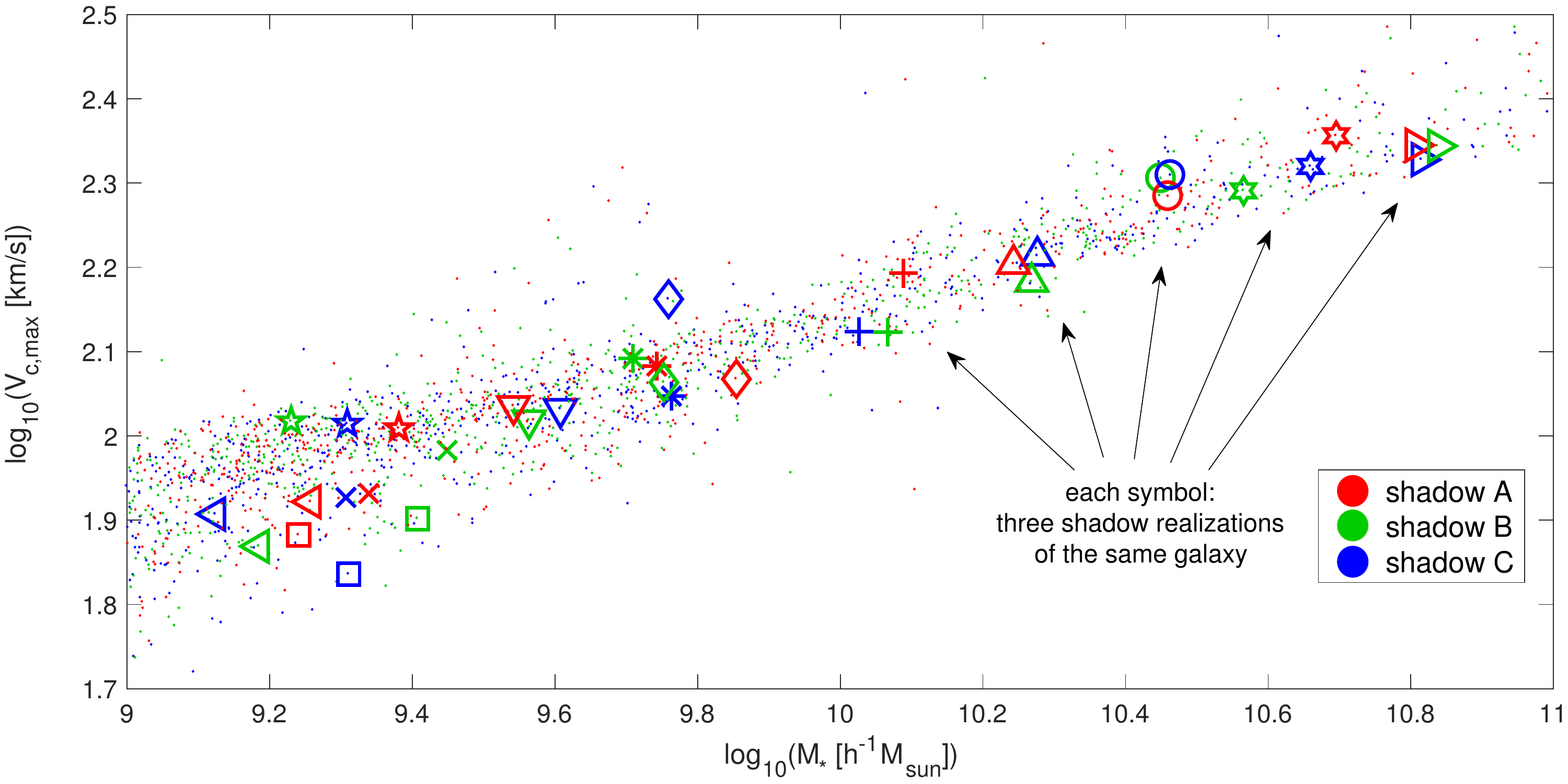}
\caption{The $z=0$ Tully-Fisher relation, defined here as $V_{c,{\rm max}}$-$M_*$, from our TNG model simulations at resolution level $\epsilon=1$. Symbols represent individual galaxies, with colors distinguishing different shadow realizations that started from slightly perturbed initial conditions at $z=5$. It is visually evident that the scatter between shadow galaxies can be non-negligible compared to the total scatter in the relation.}
\vspace{0.3cm}
\label{f:TFR}
\end{figure*}

This is demonstrated explicitly in \Fig{TFR}, which shows a scatter plot of the stellar mass and the maximum circular velocity of galaxies in the $\epsilon=1$ simulation set of the TNG-model series. The full $z=0$ galaxy population in each of the three simulations in this set is shown with small dots of a different color, clearly delineating (a version of) the well-known Tully-Fisher relation and its scatter. In addition, twelve triplets of shadow galaxies are shown using large black symbols, each with a unique symbol. Some of them (crosses, hexagrams) are displaced roughly in parallel to the overall slope of the mean scaling relation. Some, however, are displaced roughly in the perpendicular direction (asterisks, diamonds). Some do not have a strong preferred direction (triangles), while some are displaced mostly along one of the axes (pentagrams, circles). It is possible that shadow versions of certain galaxies indeed intrinsically tend to be displaced in certain preferred directions, or perhaps these particular cases are just random draws from an underlying distribution of displacements that is similar for all galaxies. To distinguish these two possibilities would require having a large number of shadow versions for a sizable number of galaxies, but since we only have a small number of shadow versions for each galaxy (albeit for a large number of galaxies), our setup does not allow us to address this specific question any further.

\begin{figure*}
\centering
\includegraphics[width=1.0\textwidth]{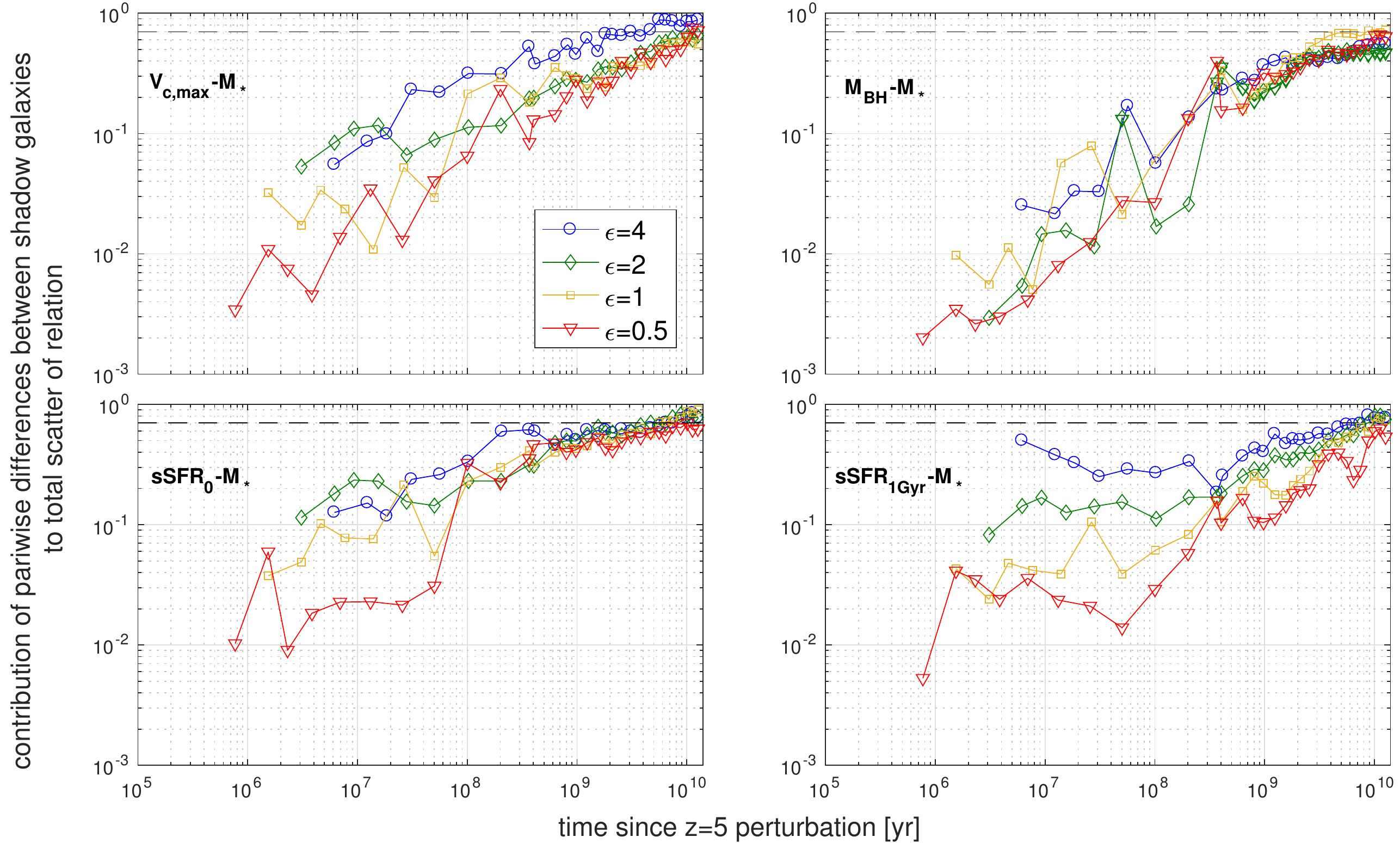}
\caption{The evolution of the fractional contribution of pairwise differences between shadow galaxies to the total scatter in various scaling relations, for galaxies with mass of $9.5<\log M_*[h^{-1}\Msun]<10$ (in the middle-right panel: $11.5<\log M_{\rm DM}[h^{-1}\Msun]<12$) in our TNG model simulation series.
Each panel presents a distinct scaling relation, as indicated in its upper-left corner, using four resolution levels, which are indicated by color, increasing from blue to red. The quantity on the vertical axis is the ratio of two quantities: in the numerator, the standard deviations of the pairwise logarithmic differences between shadow galaxies in the direction perpendicular to the respective scaling relation, divided by $\sqrt{2}$; in the denominator, the total scatter of that relation in the same, perpendicular direction. These ratios are shown as a function of time since $z=5$, when perturbations were applied. The results for these scaling relations are rather stable at a contribution of around $\sim(70\%)^2\sim50\%$ to the variance of the relations from the butterfly effect (this level is indicated with black dashed horizontal lines).}
\vspace{0.3cm}
\label{f:contribution_vs_time_TNG_converged}
\end{figure*}

\begin{figure*}
\centering
\includegraphics[width=1.0\textwidth]{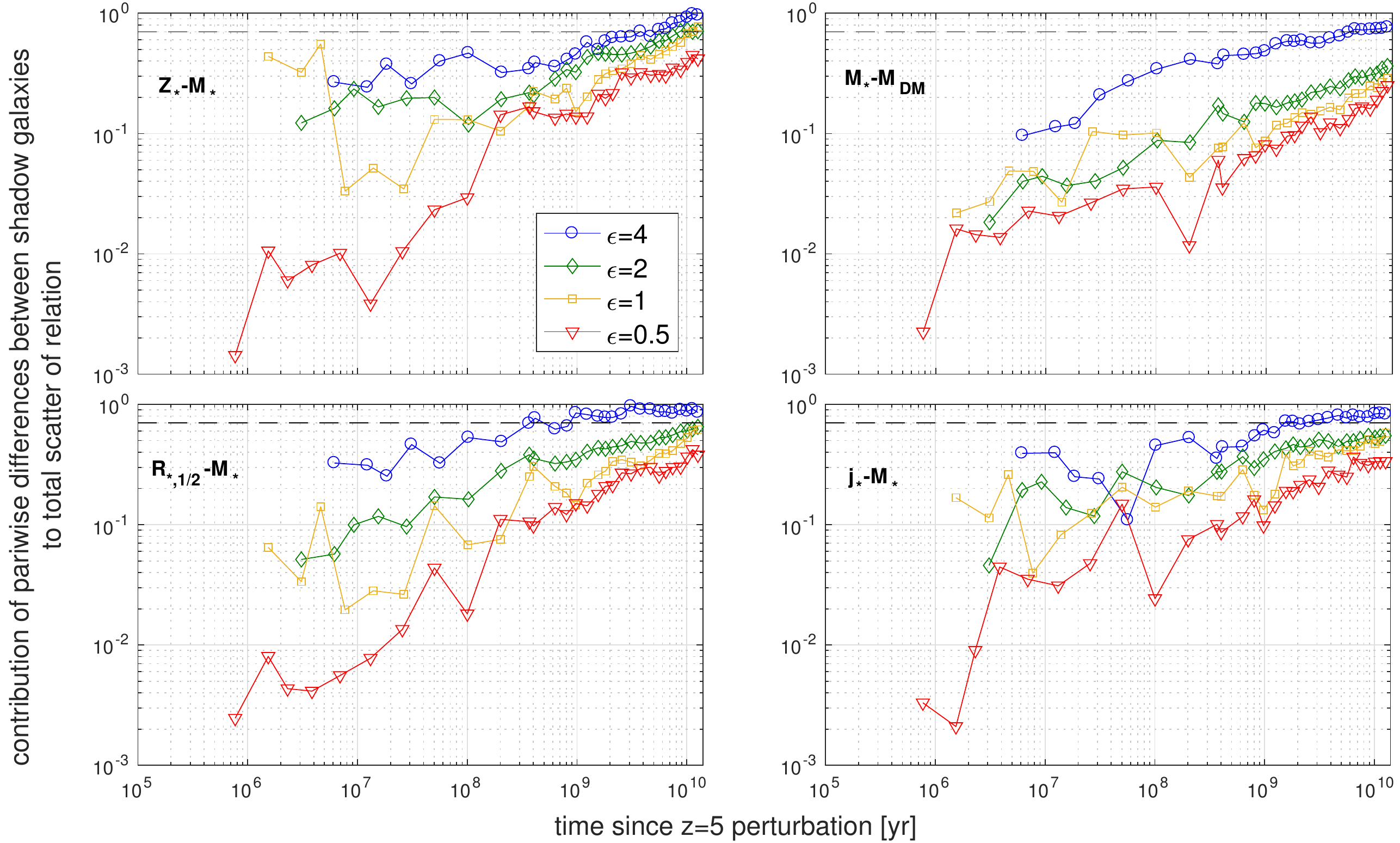}
\caption{Same as \Fig{contribution_vs_time_TNG_converged}, but for scaling relations where convergence is less clear, and at $z=0$ the relative contributions at the highest resolution are smaller, $\sim(30\%)^2\sim10\%$.}
\vspace{0.3cm}
\label{f:contribution_vs_time_TNG_unconverged}
\end{figure*}

\Fig{TFR} suggests visually that the scatter between shadow versions of individual galaxies may constitute a considerable fraction of the overall scatter in certain scaling relations in our simulations. In \Figs{contribution_vs_time_TNG_converged}{contribution_vs_time_TNG_unconverged} this notion is quantified for a selection of eight scaling relations with the TNG-model series, using the procedure described in Section \ref{s:methods}. In \Fig{contribution_vs_time_TNG_converged}, shown are the Tully-Fisher relation $V_{c,{\rm max}}$-$M_*$ (top left), the black hole mass-stellar mass relation $M_{\rm BH}$-$M_*$ (top right), and the star formation main sequence using two different timescales, ${\rm sSFR}_0$-$M_*$ (bottom left) and ${\rm sSFR}_{1\Gyr}$-$M_*$ (bottom right). Further, \Fig{contribution_vs_time_TNG_unconverged} presents the mass-metallicity relation $Z_*$-$M_*$ (top left),the baryonic conversion efficiency $M_*$-$M_{\rm DM}$ (top right), the size-mass relation $R_{*,1/2}$-$M_*$ (bottom left), and the relation between stellar specific angular momentum and stellar mass $j_*$-$M_*$ (bottom right). In particular, what is shown as a function of post-perturbation time is the ratio between the inferred standard deviations between shadow galaxies in the direction perpendicular to the various scaling relations and the standard deviations of the full galaxy population in that same direction, namely the intrinsic scatter of the relations. This can be thought of as the fractional contribution of the butterfly effect to the total scatter of the relations. More precisely, under the reasonable assumption that the butterfly effect and additional effects contribute to the scatter independently, and hence contributions should be summed in squares, the square of the quantity shown on the vertical axes is the fractional contribution of the butterfly effect to the variance of the scaling relations.

\Figs{contribution_vs_time_TNG_converged}{contribution_vs_time_TNG_unconverged} present what we regard as the central result of this work. \Fig{contribution_vs_time_TNG_converged} shows that with the TNG model the butterfly effect contributes at late cosmic epochs $\sim50\%$ of the variance (the square of the scatter) around the Tully-Fisher relation, the $M_{\rm BH}$-$M_*$ relation, and the the star formation main sequence. For the former relation, this contribution is $\gtrsim20\%$ for most of cosmic time ($z<1$), while for the latter, it is $\gtrsim40\%$ throughout this time window. These results are very convincingly converged. In contrast, the contributions of the butterfly effect to the size-mass, angular momentum-mass, baryonic conversion efficiency and stellar mass-metallicity scaling relations, shown in \Fig{contribution_vs_time_TNG_unconverged}, are much smaller and less clearly converged with increasing resolution. At our highest resolution, the contribution at late times to the variance around the these relations is $\sim10\%$. 

It is interesting to consider which of the two quantities making up each of the relations contributes more significantly to these results. The cases of the $M_{\rm BH}$-$M_*$, ${\rm sSFR}$-$M_*$ and $M_*$-$M_{\rm DM}$ relations are all similar: the relations themselves are roughly linear, the differences between shadow galaxies in the two quantities making up the relation are uncorrelated, and one of them is larger than the other, making it the dominant contribution. As can be seen in \Fig{differences_vs_time_TNG}, the differences in stellar mass are larger than those in dark matter mass, making the butterfly effect for the stellar mass dominate its relative contribution to the scatter in the $M_*$-$M_{\rm DM}$ relation. Similarly, the differences in SFR and those in black hole mass are larger than those in stellar mass, making the former two dominate the total contribution of the butterfly effect to the scatter in the star formation main sequence and the $M_{\rm BH}$-$M_*$ relation, respectively. For the size-mass and mass-metallicity relations, the picture is slightly different, since the relations themselves are not linear but sub-linear. This means that differences in stellar mass contribute less significantly to the scatter in these relations than equal differences in size or metallicity, since those in stellar mass displace galaxies more parallel to the relation than perpendicular to it. The implication of this is that the differences in size (metallicity) dominate the butterfly effect contribution to the scatter in the size-mass (mass-metallicity) relation. All these results hold true at all the resolution levels we probed.

The case of the Tully-Fisher relation is most involved in this respect. At low resolution, the differences in stellar mass are an order of magnitude larger than those in maximum circular velocity (as can be seen in \Fig{differences_vs_time_TNG}), hence in spite of the flatness of the $V_{c,{\rm max}}$-$M_*$ relation, the differences in stellar mass dominate the contribution to the overall scatter in the relation. This is however driven by the non-convergence of the stellar mass differences at the $\epsilon=4$ resolution level. At the highest resolution, in comparison, the differences in $M_*$ are smaller, while the differences in $V_{c,{\rm max}}$ are similar, rendering the latter the dominant contributor to the overall butterfly effect contribution to the scatter in the relation. Note that it is still the case even at the highest resolution that the differences in $V_{c,{\rm max}}$ are smaller than those in $M_*$, but since the $V_{c,{\rm max}}$-$M_*$ relation is sub-linear, the former are nevertheless dominant in the total scatter of the relation.

\section{Summary and Discussion}
\label{s:summary}

\subsection{Summary}
\label{s:summary_sub}

In this paper we investigate the response of cosmological simulations, in particular hydrodynamical ones that include models for galaxy formation, to minute perturbations to their initial conditions. The main metrics we use are global, integrated properties of galaxies such as their mass, peak circular velocity, or star-formation rate, and our samples contain hundreds to thousands of galaxies since our simulations are of uniform-resolution cosmological boxes. We find that minute differences, close to the machine precision, that we introduce between sets of otherwise identical `shadow' simulations at early cosmic times grow over billions of years by many orders of magnitude. We hence determine that `the butterfly effect' is present in our cosmological hydrodynamical simulations. To understand whether the magnitude of the effect is large enough to be of general interest for galaxy formation theory, we quantify the typical uncertainty on various simulated galaxy properties that the effect induces, and moreover quantify the contribution of the effect to the scatter in various galaxy scaling relations. Before further discussing our results as well as their relation to the real universe, we summarize them as follows.

\begin{itemize}
\item \Figs{differences_vs_time_SP}{differences_vs_time_TNG}: The divergence rate between shadow simulations, and in particular the existence of a saturation level and its magnitude, is not universal but varies with the considered quantity, the physics included in the simulation, and numerical resolution. Generally speaking, the resolution dependence of the results is much weaker in simulations that include stellar and black hole feedback than in those that include no feedback. This implies that at the highest resolution we consider, which is better than that of the Illustris and TNG100 simulations, differences between shadow simulations that include feedback are larger than between those that do not.
\item \Fig{differences_vs_time_TNG}: After $\sim10\Gyr$ of cosmic evolution, the differences between shadow simulations that utilize our fiducial feedback model, in terms of all the baryonic galaxy properties that we explore, are still growing. At our highest resolution, by $z=0$ they reach a level of $\sim0.01\dex$ (namely, a few percents) for peak circular velocity, $\sim0.1\dex$ (namely, tens of percents) for stellar half-mass size, star-formation rate, black hole mass and angle between halo and galaxy angular momentum vectors, and in between those values for stellar mass and stellar metallicity. The differences of dark matter mass, on the other hand, have already reached a constant level of $\sim0.01\dex$ after $\sim1\Gyr$ of evolution.
\item \Fig{no_rand_sims}: Given enough time to evolve, the results are largely robust to whether random numbers are used in the subgrid models, as is standard in cosmological simulations, or whether their usage is completely avoided. Appendix \ref{s:verification} shows that when random numbers are avoided, whether in DM-only or hydrodynamical simulations, the initial growth of the perturbations is approximately exponential with a timescale on the order of the dynamical time of the relevant systems (dark matter halos or galaxies, respectively). This is the behavior expected from a chaotic system. Later on, the evolution slows down into a power-law growth regime.
\item \Figs{2d_hists}{TFR}: On a galaxy-by-galaxy basis, the differences between shadow galaxies in the values of different properties are largely uncorrelated. This means that for the scaling relations between two distinct galaxy properties that we examined, for example the Tully-Fisher relation, the separations between sets of shadow galaxies are sometimes roughly aligned with the relation but sometimes are roughly perpendicular to it. In other words, the scatter about (i.e.~perpendicular to) scaling relations arises not only due to macroscopic differences in initial conditions between different galaxies, which determine e.g.~the large-scale tidal field and the timings and mass ratios of mergers in their formation history, but also due to the sensitivity of the final galaxy properties to the `microscopic' initial conditions.
\item \Figs{contribution_vs_time_TNG_converged}{contribution_vs_time_TNG_unconverged}: Quantifying the previous point, we find that the scatter perpendicular to the Tully-Fisher relation between shadow galaxies with $9.5<\log M_*[h^{-1}\Msun]<10$ reaches, at late cosmic times, a value that is approximately $70\%$ of the total scatter in the relation. This means that about one half of the variance around the mean relation arises from the chaotic-like behavior of the simulation. Similar or even higher values are found for the sequence of star-forming galaxies between their SFR and their stellar mass and for the relation between black hole mass and host galaxy stellar mass. In contrast, for the relations between stellar mass and halo mass as well as between stellar size, angular momentum or metallicity and the stellar mass, the contribution of the butterfly effect to the overall relation scatter is not converged and is lower at higher resolutions. In particular, at our highest resolution, the butterfly effect only contributed a few percents of the variance about these relations.
\end{itemize}

\subsection{Implications for Interpretation of Simulations}
\label{s:discussion_simulations}

First, we should emphasize that, in principle, as a result of sample variance, the effect we have studied -- namely differences between shadow galaxies -- propagates into differences of the properties of the ensemble of galaxies between shadow simulations. This means that ensemble properties, such as the mean and scatter of scaling relations between two quantities or the total stellar mass or SFR in the simulation box, may differ between two shadow simulations simply because each and every galaxy is different to a certain extent from its shadow. However, we consider this effect to be unimportant in most cases, since ensemble differences shrink toward zero as the number of galaxies increases, due to the central limit theorem. In other words, if the simulation volume is small and contains only a small number of galaxies, ensemble properties of those galaxies will be sensitive to the individual galaxies; instead, for larger and larger number of galaxies in the ensemble, the differences between the shadows will tend to cancel out more and more completely, leaving the average statistical properties of the ensemble of galaxies less and less affected. Nevertheless, in regimes in which a small number of galaxies is considered, for example by applying some cuts in a multi-dimensional parameter space, the ensemble properties may be strongly enough affected by the uncertainty of the properties of the individual galaxies the ensemble is comprised of. For example, the stellar mass function at the highest-mass end of any simulation box is by definition based on a small number of the most massive galaxies in the simulation. The uncertainty on those masses implied by the butterfly effect (e.g.~at a level of a few percent, \Fig{vs_mass_TNG}) will then translate to a similar level of horizontal uncertainty of the mass function itself. This in turn, if the mass function is steep, will translate into a larger vertical uncertainty, which may be needed to be taken into consideration.

A distinct implication of this work pertains to our ability to explain the scatter in scaling relations or in galaxy properties using deterministic considerations. If for given initial conditions and a given physical model, each galaxy may occupy a finite rather than infinitesimal region in property space, then its properties can only partially be predicted based on its initial conditions and a set of physical arguments, or in other words, there is a limit to the degree to which one can `understand' the properties of that galaxy. When applied to a galaxy population, this kind of argument implies that only a fraction of the scatter in galaxy properties or in correlations between them can be understood, and once a correct model explains that fraction, the understanding is in fact complete. If the butterfly effect exists in real galaxy formation as it does in our simulations (a possibility discussed below), then these arguments apply to the scatter and scaling relations of galaxies in the real universe.

If, however, the effect we measure in the simulations does not exist or is much larger than in the real universe, then the implication may be that some of the simulated scatter is artificially inflated by the numerics. In this case, care should be taken when comparing the simulated scatter to the observationally-inferred one. For example, if the simulated scatter is smaller than the intrinsic scatter inferred from observations (as has been argued to be the case for the Tully-Fisher relation, e.g.~\citealp{McGaughS_12a}, and as is probably the case for the black hole-stellar mass relation, e.g.~\citealp{WeinbergerR_17a}), then the tension between the two may in fact be even starker than it might appear without considering the numerical butterfly effect. Conversely, if the simulated scatter is larger than observed, the numerical butterfly effect could account for the discrepancy.

In these considerations we have implicitly assumed that scatter driven by the butterfly effect is independent of, and can be added e.g.~in quadrature, to other sources of scatter, namely scatter arising from `macroscopic' differences between the environments and initial conditions of different galaxies. However, this does not necessarily have to be the case. If all the butterfly effect did was `shuffling' galaxy properties between different galaxies, the methods used in this work would detect a non-zero butterfly effect, while the overall scatter between galaxies would not increase.

A more specific scenario where such a situation could arise is one where the scatter between galaxies is associated with short timescale oscillations of galaxy properties. These oscillations could be driven by some physical process regardless of the butterfly effect. In this case, the butterfly effect can be thought of as merely determining the `phase' of each galaxy within the oscillation pattern, but as the driver of neither the pattern itself nor of the scatter of galaxy properties associated with it. In such a scenario, the evolution paths of shadow galaxies in some physical property space will be oscillating around some mean path and may be recurrently crossing each other (as opposed to monotonically drifting away from each other). In this case, measuring the butterfly effect on time-averaged galaxy properties will result in a diminished effect compared to instantaneous properties. We leave further considerations along these lines to future work, but point out that in the single case we have examined using a time-averaged measurement, namely that of ${\rm SFR}_{1\Gyr}$, the magnitude of the butterfly effect we found was very close to that of the instantaneous property ${\rm SFR}_0$. Note that even in the oscillatory scenario, our measurement of the magnitude of the butterfly effect is an indication of the level to which the properties of individual galaxies can(not) be predicted from first principles, even if the level of scatter between galaxies can be attributed to the physical processes driving the oscillations rather than to the butterfly effect itself.

Our work also has implications for the interpretation of differences between simulations with different physical models or numerical schemes on an object-by-object basis, most notably in the context of `zoom-in' simulations. In order to conclude that the properties of a certain simulated galaxy differs between two simulations due to changes to the numerical scheme (including thereby both physical processes and their particular implementation), it first must be determined that these differences do not arise due to the butterfly effect alone. Unless a large ensemble of shadow simulations is available, which is normally not the case, this implies that the changes have to be significantly larger than the typical magnitude of the butterfly effect in order to be considered `real'. A complication that arises is that the magnitude of the butterfly effect on the considered quantity it is not a-priori known, since as we have demonstrated here, that magnitude itself varies with the physical model as well as with numerical resolution. That \citet{KellerB_18a} found an opposite effect of feedback to the one we found, namely that in their simulations feedback acts to reduce the magnitude of the butterfly effect rather than enhance it as in ours, serves as further evidence that the dependence on physical and numerical approach may be significant, and is complicated.

\subsection{Implications for Galaxy Formation in the Real Universe}
\label{s:discussion_reality}

A fundamental question that underlies the work presented here is whether the effect we have identified is purely numerical, i.e., applicable only to the simulated systems, or physical, in the sense that it exists in the real universe as well. We do not have a clear answer to this question, but we discuss several interesting aspects of it.

First, one can ask whether the inherently limited accuracy of the integration of the gravity and hydrodynamics equations may be introducing chaos into the system. For example, an infinitesimal change in the position of a dark matter particle or a gas mesh-generating point may result in a finite change to the forces or the fluxes due to a finite change in the structure of the gravity tree or the geometry of the mesh. This clearly results in the amplification of some differences, at least locally and on short timescales. The fundamental question is however whether it is this kind of amplification that builds up gradually toward the macroscopic effect we have quantified, or whether those purely numerical effects tend to cancel out.

A second, related question is whether the use of probabilistic modeling in the simulations (which cannot be trivially converted into continuous/non-probabilistic formulations) introduces chaos into the numerical system that does not exist in the physical reality. The probabilistic algorithmic implementation uses random number generators to control several physical processes. We find that infinitesimal changes in our simulations may result in discrete changes of finite magnitude within a single simulation time step due to changes in the `field' of random numbers as a function of space and time. Indeed, we show in Appendix \ref{s:verification_TNG} that differences between shadow simulations grow faster once their respective random number sequences are effectively no longer the same.

We believe that it is at least plausible that, even if these numerical drivers of the butterfly effect exist only in the simulations but not in physical reality, our results still largely apply to galaxy formation in the real universe as well, due to physical drivers of the butterfly effect. This is because galaxies contain chaotic systems of various natures and scales, which inject chaos into the galactic scale in analogy with the purely numerical factors described above. For example, even in a purely gravitational system without any discrete effects in the force calculation, some satellite galaxies and stars are on truly chaotic orbits within their dark matter halos. Further, if turbulence in molecular clouds is truly chaotic (e.g.~\citealp{DeisslerR_86a,BohrT_05a}) then chaos determines where and when individual stars form and hence where and when they explode.
These are `discrete' events that are analogs to the choice of a random number to determine, e.g., the birth time and place of a star in the unresolved interstellar medium in our simulations. In some aspects our simulations are most likely to actually suppress chaos that exists in reality. For example, the flow in the interstellar medium in our simulations is less turbulent than in reality due to numerical viscosity. Another example is the lack of stochasticity in the sampling of the initial mass function (IMF) in our simulations, in which each stellar population is comprised of a `smooth', idealized IMF.

While the nature of chaos injection from small scales into the galactic scale differs between our simulations and reality, as discussed, it is possible that the growth of differences in macroscopic galaxy properties that is exhibited by our simulations captures a real phenomenon. This is the third, `dynamical', phase discussed at the end of Appendix \ref{s:verification_TNG}, during which the growth of differences is no longer exponential but instead power-law or slower, but during which most of the growth in absolute terms is achieved. It is instructive in this context that we find a significant difference in the characteristics of the butterfly effect between our simulations with and without feedback. In spite of having the same small-scale chaos drivers such as roundoff errors, discreteness effects and random numbers as the simulations without feedback, those with feedback result in a stronger butterfly effect. This suggests that it is the nature of the dynamics on galactic scales that determines the degree to which `random' differences e.g.~in the formation sites of stars develop into global differences in galaxy properties.

Even under the assumption that this is indeed the case, our work nevertheless cannot yet determine with great certainty what is the magnitude of the butterfly effect on galaxy formation in the real universe. Additional work would be required in order to characterize and understand the dependence of this magnitude on the physical models used in the simulation. It is possible that eventually only an accurate simulation of galaxy formation, perhaps much more accurate than ours, will be reliable enough to parallel the real universe in this respect.

\acknowledgements
We thank Paul Torrey for comments on the manuscript.
The simulations analyzed in this study were run on the Iron cluster at the Flatiron Institute and the Gordon-Simons system at the San Diego Supercomputer Center, as well as on the Stampede supercomputer at the Texas Advanced Computing Center through XSEDE allocation AST160026. We are grateful to the scientific computing teams at all three of these institutions for their continual and dedicated technical assistance and support. The Flatiron Institute is supported by the Simons Foundation.
GLB acknowledges support from NSF grant NNX15AB20G and NSF grant AST-16-15955.
RW acknowledges support through the European Research Council under ERCStG grant EXAGAL-308037, and would also like to thank the Klaus Tschira Foundation.

\appendix

\section{Dark Matter-only Simulations}
\label{s:DMonly}

\begin{figure*}
\centering
\includegraphics[width=1.0\textwidth]{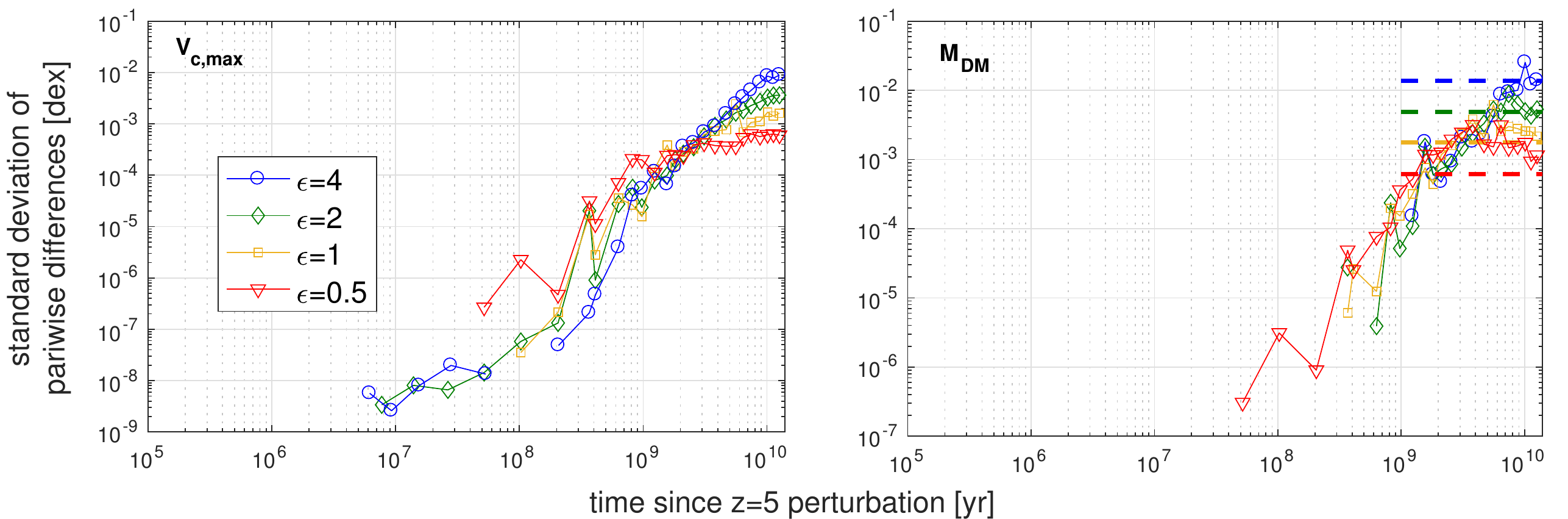}
\caption{The evolution of differences between shadow galaxies with final halo mass of $11.5<\log M_{\rm DM}[h^{-1}\Msun]<12$ in our dark matter-only simulation series, similarly to \Figs{differences_vs_time_SP}{differences_vs_time_TNG} (note however the different scale on the vertical axes). Two physical quantities are shown: maximum circular velocity (left) and halo mass (right), each based on four resolution levels, which are indicated by color, increasing from blue to red.}
\vspace{0.3cm}
\label{f:differences_vs_time_DM}
\end{figure*}

In \Fig{differences_vs_time_DM} we show that the initial minute perturbations we apply to particle positions at $z=5$ evolve into percent-level differences in halo properties even in the DM-only case. At early times the various resolution levels evolve similarly. At late times, however, lower resolution levels show continuously increasing differences, while higher resolution levels show a weaker growth rate, which for the highest level, $\epsilon=0.5$, appears already as a plateau at $t\gtrsim2\Gyr$. For mass (right panel), this plateau level is however still very close to the one expected just from shot noise given the finite number of dark matter particles in the halos. Hence, the four resolution levels we have are not enough to clearly determine whether the result is converged or will continue shrinking with even higher resolution. This is different from the cases with hydrodynamics and galaxy formation models discussed in the main part of the paper.

\section{Special simulations for numerical verification}
\label{s:verification}

This Appendix has two main goals: explore the sensitivity of the results to several numerical nuisance parameters, both for pure N-body cosmological simulations containing only dark matter and for hydrodynamical runs, and support the interpretation of our main results with regards to the role of the usage of random numbers in the baryonic physics models.

Table \ref{t:verification} provides an overview of the additional sets of simulations shown in the figures of this Appendix. The special features distinguishing these simulations from the fiducial ones fall into three categories: {\bf (i)} greater numerical integration accuracy through the usage of smaller simulation time steps, {\bf (ii)} variations of the magnitudes of the initial perturbations applied to the $z=5$ initial conditions of the shadow simulations, and {\bf (iii)} a different usage of random numbers. We run several sets of DM-only verification simulations at resolution level $\epsilon=2$ and several with the TNG model at level $\epsilon=1$. In addition, there are three $\epsilon=2$ sets that are included in Table \ref{t:verification} but whose results are not shown in the figures, as they are for any practical purpose indistinguishable from the fiducial case. This includes a DM-only set with a higher accuracy in the tree part of the gravity force calculation, a TNG model set with larger initial perturbations, and a TNG model set where compilation optimization has been turned off (rendering our results insensitive to the optimization level).

\subsection{The Case of Pure Dark Matter Simulations}
\label{s:verification_DM}

\begin{table*}
\begin{tabular*}{0.99\textwidth}{@{\extracolsep{\fill}}|p{7.5cm}|c|c|c|}
\hline
Simulation type & Physics model & Resolution level & Line style \\
\noalign{\vskip 0.5mm}
\hline
\hline
\noalign{\vskip 0.5mm}
$10$x smaller simulation time step $\Delta t$ (individually) & DM-only & $\epsilon=2$ & black asterisks, \Figs{DMtests_Vmax}{DMtests_distancemetric} \\
\hline
$10$x smaller gravity tree opening angle & DM-only & $\epsilon=2$ & - \\
\hline
$50$x smaller maximum simulation time step $\Delta t$ (globally) & DM-only & $\epsilon=2$ & magenta triangles, \Figs{DMtests_Vmax}{DMtests_distancemetric} \\
\hline
$500$x smaller maximum simulation time step $\Delta t$ (globally) & DM-only & $\epsilon=2$ & magenta dots, \Figs{DMtests_Vmax}{DMtests_distancemetric} \\
\hline
$50$x smaller maximum simulation time step $\Delta t$ (globally), and only a single particle is initially perturbed & DM-only & $\epsilon=2$ & red circles, \Figs{DMtests_Vmax}{DMtests_distancemetric} \\
\hline
$10^7$x larger initial perturbation & DM-only & $\epsilon=2$ & cyan squares, \Figs{DMtests_Vmax}{DMtests_distancemetric} \\
\hline
$10^7$x larger initial perturbation & TNG model & $\epsilon=2$ & - \\
\hline
No code optimization (compilation with \texttt{-O0}) & TNG model & $\epsilon=2$ & - \\
\hline
$50$x smaller maximum simulation time step $\Delta t$ (globally) & TNG model & $\epsilon=1$ & magenta triangles, \Fig{TNGtests} \\
\hline
$50$x smaller maximum simulation time step $\Delta t$ (globally) and different usage of random numbers, method 1 & TNG model & $\epsilon=1$ & purple pentagrams, \Fig{TNGtests} \\
\hline
$500$x smaller maximum simulation time step $\Delta t$ (globally) and different usage of random numbers, method 2 & TNG model & $\epsilon=1$ & gray crosses, \Fig{TNGtests} \\
\hline
$500$x smaller maximum simulation time step $\Delta t$ (globally) and no usage of random numbers & TNG model & $\epsilon=1$ & dark blue lines, \Fig{TNGtests} \\
\hline
\end{tabular*}
\caption{\small {\bf An overview of the numerical verification simulations.} We have generated six numerical verification pairs of shadow DM-only simulations at resolution level $\epsilon=2$, five of which are shown in \Figs{DMtests_Vmax}{DMtests_distancemetric}. Each of these pairs has a unique difference in its setup with respect to the fiducial simulations discussed throughout the paper, as briefly summarized in the first column and discussed in detail in the Appendix. Two distinct $\epsilon=2$ pairs were run with the TNG model, which produce virtually indistinguishable results to the fiducial case and are therefore not explicitly shown. Four numerical verification pairs were run at resolution level $\epsilon=1$ with the TNG model, presented in \Fig{TNGtests}.}
\label{t:verification}
\end{table*}

\begin{figure*}
\centering
\includegraphics[width=1.0\textwidth]{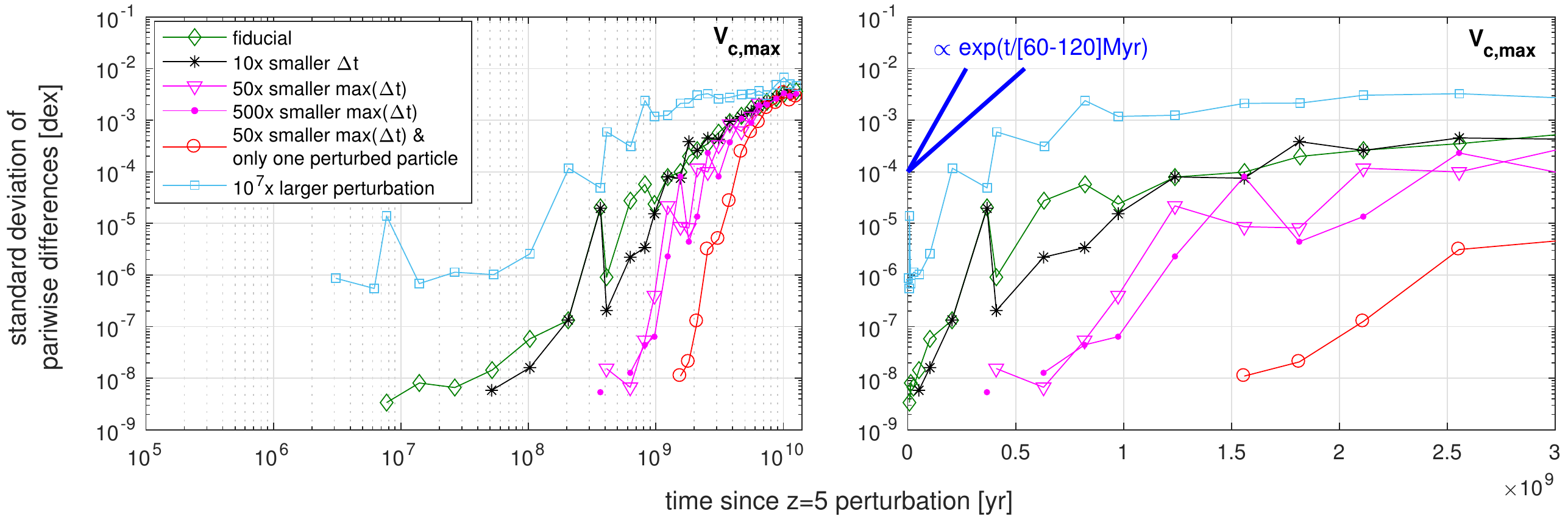}
\caption{Growth of the standard deviations of the distributions of pairwise differences between the $V_{c,{\rm max}}$ of shadow galaxies in various tests of DM-only simulations at resolution level $\epsilon=2$, as indicated in the legend (see also Table \ref{t:verification}). Both panels show the same data, focusing on different timescales. Note that the vertical axis spans a significantly larger dynamic range than the analogous \Figs{differences_vs_time_SP}{differences_vs_time_TNG}. {\it Right:} The first $3\Gyr$ on a linear time axis. Non-zero differences appear for the various sets at different times, but then they all exhibit an initial growth phase that is close to exponential with a Lyapunov exponent of $\sim1/100\Myr$, as indicated by the thick blue lines in the top-left corner. After this initial phase that lasts $\sim1\Gyr$, the growth slows down. {\it Left:} The full cosmic time on a logarithmic time axis. The initial exponential growth phase turns roughly to a power-law, and all set converge to essentially the same outcome by $z=0$ in spite of the vastly different result at earlier times such as $t=1\Gyr$ or $t=2\Gyr$.}
\vspace{0.3cm}
\label{f:DMtests_Vmax}
\end{figure*}

\Fig{DMtests_Vmax} presents the standard deviations of the $V_{c,{\rm max}}$ differences distributions between shadow subhalos in DM-only simulations, similarly to \Fig{differences_vs_time_DM}, but for the numerical verification sets. The left panel shows the usual log-log view, while the right panel shows linear time on the horizontal axis, and only up to $t=3\Gyr$. It becomes clear from examination of the right panel that at early times, $t\lesssim0.5\Gyr$, the evolution can be well fit by an exponential growth of the differences in time. This is a characteristic of chaotic systems, and the thick blue lines in the top-left of that panel indicate exponential growth rates with Lyapunov timescales of $60\Myr$ and $120\Myr$, roughly bracketing the slopes seen for the various cases in their initial phases. It is worth noting that given that the dynamical time of dark matter halos (defined as $10\%$ of the contemporaneous Hubble time) at the time the perturbations are applied is $117\Myr$, these Lyapunov timescales are not surprising  \citep{KandrupH_91a,GoodmanJ_93a,KandrupH_94a}.

The fiducial case (at the $\epsilon=2$ level, as all other simulations in \Fig{DMtests_Vmax}) is shown in green, and on the left panel is identical to the line with the same style in \Fig{differences_vs_time_DM}.
One of the special cases, where the simulation time step was decreased uniformly by a factor of $10$ for all particles (black asterisks), appears to behave essentially just like the fiducial case. The same holds for an additional case, which is listed in Table \ref{t:verification} but not shown in \Fig{DMtests_Vmax} for visual clarity, where the force calculation accuracy of the tree algorithm was significantly increased by decreasing the node opening angle threshold \citep{HernquistL_87a}. However, when forcing the time steps of {\it all} particles to a common, smaller maximum time step (magenta triangles and dots for factors of $50$ and $500$ compared to the fiducial simulations, respectively), then the results are affected. In particular, a very similar evolution occurs, namely exponential with a similar timescale, but it is delayed with respect to the fiducial case. The origin of this behavior will be elucidated when discussing the next figure.

Before doing so, we point out the two sets where the nature of the initial perturbation has been modified. In one case (cyan squares), each particle is displaced initially at its eighth significant digit instead of the fifteenth, namely by up to $\approx1\pc$ instead of $10^{-7}\pc$ (corresponding roughly to a `single precision' perturbation). This results in $V_{c,{\rm max}}$ differences that are initially about two orders of magnitude larger than in the fiducial case, but the initial growth is still exponential with a similar timescale, such that a plateau is reached earlier -- but to the same level as in the fiducial case. In the second case (red circles), the initial perturbation (at the fiducial magnitude) is applied only to a single particle in the whole simulation box. In this case, non-zero $V_{c,{\rm max}}$ differences take $1.5\Gyr$ to appear, but thereafter evolve similarly. After several more billions of years, they again reach the same level as the fiducial case. Next we discuss the nature of this delay as well.

\begin{figure*}
\centering
\includegraphics[width=1.0\textwidth]{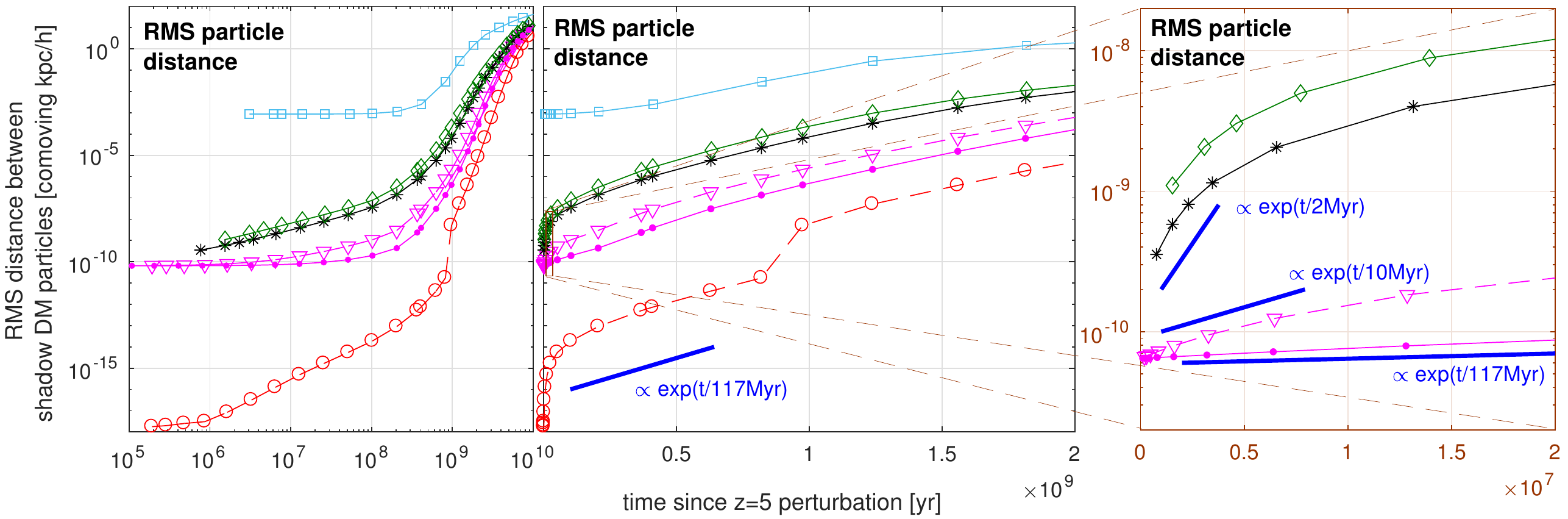}
\caption{Growth of the root-mean-square physical distances between shadow dark matter particles in the various $\epsilon=2$ DM-only sets: the fiducial one (green) as well as various tests (line styles as in \Fig{DMtests_Vmax}). All panels include the same data, but focus on different scales. {\it Left:} The full range of RMS distances with a logarithmic time axis. While the initial perturbations span $15$ orders of magnitude, by $z=0$ they all converge to a very similar outcome. At $t\gtrsim1\Gyr$, the growth is approximately a power-law with time. {\it Middle:} The full range of RMS distances with a linear time axis limited to the first $2\Gyr$. The growth during most of the first $\Gyr$ is roughly exponential with a Lyapunov exponent of $\sim1/117\Myr$, which is the dynamical time of dark matter halos at $z=5$. At $t\gtrsim1\Gyr$, the growth starts slowing down toward the power-law behavior seen in the left panel. {\it Right:} A zoom-in on the first $20\Myr$, focusing on the simulations with the fiducial type of perturbation. The set using very small time steps (magenta dots) shows the exponential with a timescale of $\sim117\Myr$ right from the beginning, while sets using larger time steps show an initial phase of faster, numerical error-driven, growth.}
\vspace{0.3cm}
\label{f:DMtests_distancemetric}
\end{figure*}

In \Fig{DMtests_distancemetric} we present the time evolution of a different kind of quantity from the ones discussed so far. In this case, it is the root-mean-square (RMS) of the distances between the positions of matched individual dark matter particles between pairs of shadow simulations. This is a useful quantity because unlike differences in global subhalo properties, these distances are continuous and are directly related to the initial perturbation, which is implemented as a displacement of particle positions. Indeed, the left panel shows that the cases with the fiducial kind of perturbation begin at a level of $10^{-10}\hkpc$, as prescribed, and the case with the larger initial perturbation is correspondingly at $10^{-3}\hkpc$. Finally, the case where only a single particle is perturbed appears initially at a level of $10^{-18}\hkpc$, which is indeed what is expected given that the number of particles in our DM-only $\epsilon=2$ simulations is $512^3\approx1.3\times10^{8}$.

The left panel of \Fig{DMtests_distancemetric} shows the usual log-log view, while the middle and right panels show time on a linear axis, up to $t=2\times10^9\Gyr$ in the former, and further zooming in on $t\leq2\times10^7\Gyr$ in the latter, which also focuses on a particular range on the vertical axis (as symbolized by long dashed brown lines) that includes only the simulations with the fiducial kind of perturbations. In the middle panel it is seen that all simulation types show in the first $\sim\Gyr$ of evolution an exponential growth of the RMS distance between shadow particles, with a Lyapunov time consistent with the dynamical time of dark matter halos at $z=5$ ($117\Myr$), as indicated by the thick blue line at the bottom of the panel\footnote{We have confirmed that it is indeed dark matter particles that are part of dark matter halos that drive the growth, rather than particles outside of halos, see also \citet{ThiebautJ_08a}.}. This is very similar to the evolution of the $V_{c,{\rm max}}$ differences seen in \Fig{DMtests_Vmax}. The transition to an approximate power-law growth at $t\gtrsim1\Gyr$ is also similar, as is the convergence by $z=0$ to a very similar result in all cases despite their vastly different initial stages and evolutions\footnote{We note that the initial imposed correlation between the magnitude of the perturbation and the position in the box is gradually erased over time, as the results converge toward a value that is independent of the magnitude of the initial perturbation (this happens even faster in the TNG model case than with DM only). This is analogous to the shrinking differences between simulations with different overall magnitudes of initial perturbations.}.

However, a close examination of the first $20\Myr$ in the right panel of \Fig{DMtests_distancemetric} demonstrates that in the very initial phase, the `effective' Lyapunov time actually differs between the different simulations. In particular, the growth in the fiducial case (green) in the first few million years has no exponential form altogether, but instead can be seen in the left panel to be a power-law with an index of unity. The `effective' growth timescales when exponential fits are forced, indicated with blue thick lines and associated timescales, become longer as the maximum simulation time step is reduced. Importantly, in the most aggressive case (magenta dots), the growth timescale appears to converge to the $\sim117\Myr$ level that then continues throughout the first $\sim1-2\Gyr$, as seen in the middle panel. We interpret these results to imply that the fiducial case is affected by numerical accuracy errors that present themselves as an early power-law growth with an artificially short associated growth timescale, but that these can be mitigated by using shorter time steps, in which case the evolution from the very beginning is physical and is on the expected, dynamical timescale\footnote{In the case where only a single particle is perturbed, the growth rate of the RMS distance before the exponential growth sets in is a power law versus time with an index of $2$. The same power law holds for the {\it number of particles} with a non-zero distance between the shadows (not shown). A possible interpretation of this growth rate is that it corresponds to the growth of the volume of a perturbation in the spherical collapse model \citep{GunnJ_72a}.}.

The lack of an initial artificially fast growth rate in the simulations that use aggressively short time steps, seen clearly in the right panel, results in an effective delay with respect to the other simulations, which is clearly seen in the middle and left panels. The intermediate cases too (magenta triangles and black asterisks) are also delayed with respect to the fiducial case, by an intermediate amount\footnote{This is more significant and seen clearly in the case of the factor $50$ smaller maximum time step (magenta triangles), where the initial growth rate is intermediate too ($\sim10\Myr$). In the case of individually smaller time steps by a factor of $10$ (black asterisks), the generation of the `delay' is unresolved by the snapshot time separations we have available, i.e.~it occurs at $t\lesssim1-2\Myr$.}. This brings us back to the delay in the appearance of $V_{c,{\rm max}}$ differences that we observed in \Fig{DMtests_Vmax}. Simulations with smaller time steps first show non-zero $V_{c,{\rm max}}$ differences at a later time because it takes them longer to develop substantial RMS particle position differences, which are necessary to give rise to $V_{c,{\rm max}}$ differences. A similar case applies for the large delay of the simulation where only a single particle is perturbed. In particular, we observe that it is common to all simulations that non-zero $V_{c,{\rm max}}$ differences appear once and only once their RMS particle differences reach a level of $\sim10^{-8}-10^{-7}\hkpc$, which occurs at different times in the various cases.

We conclude that as long as the initial perturbation of positions is large enough that it has enough time to grow to a level of $\sim10^{-8}-10^{-7}\hkpc$, given a growth rate that is roughly $10\%$ of the Hubble time, then it is large enough to develop into macroscopic (percent-level) differences in global subhalo properties such as $V_{c,{\rm max}}$.

\subsection{The Case of Hydrodynamical Simulations with the TNG Model}
\label{s:verification_TNG}

\begin{figure*}
\centering
\subfigure[$V_{c,{\rm max}}$ pairwise differences]{
          \label{f:TNGtests_Vmax}
          \includegraphics[width=1.0\textwidth]{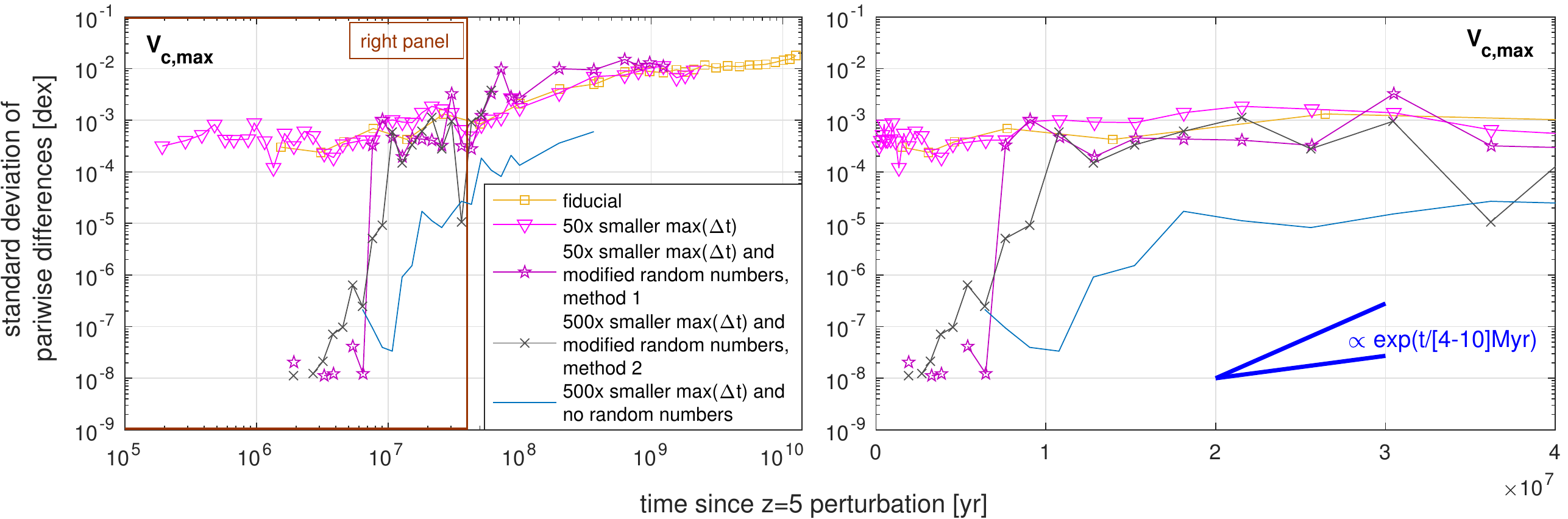}}
\subfigure[root-mean-square particle position differences]{
          \label{f:TNGtests_distancemetric}
          \includegraphics[width=1.0\textwidth]{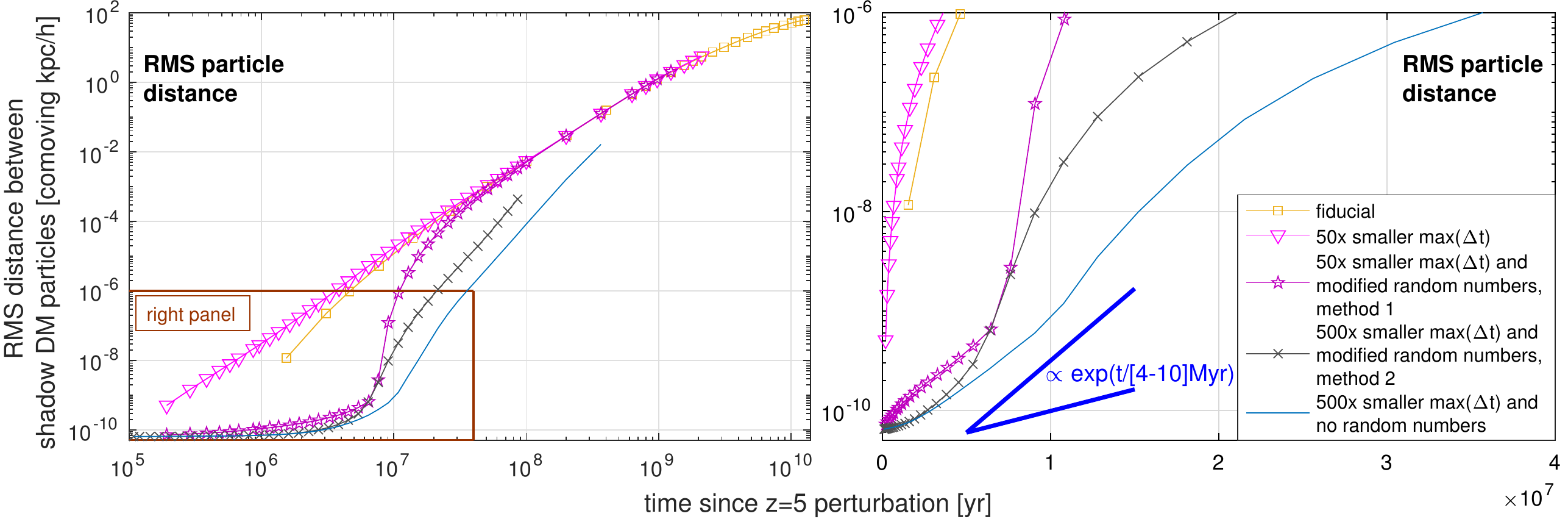}}
\caption{{\it Top:} Evolution of the standard deviations of pairwise $V_{c,{\rm max}}$ differences between shadow galaxies in the numerical verification sets of the TNG model at resolution level $\epsilon=1$ (see Table \ref{t:verification}). As opposed to the DM-only case seen in \Fig{DMtests_Vmax}, the fiducial case and the case with smaller time steps behave virtually in the same way, both showing `large' ($>10^{-4}\dex$) differences as early as can be measured ($t\approx10^6\yr$ and $2\times10^5\yr$, respectively). The cases where the random number usage is modified toward greater correspondence between the shadow simulations, or altogether removed, result in a delayed growth of differences, converging eventually toward the fiducial case. {\it Bottom:} RMS differences of the positions of shadow dark matter particles in the same sets of simulation. The fiducial and small time steps cases show a power-law growth from the very beginning, while the cases with the modified random number treatments show an initial exponential growth with a short timescale of $\sim4\Myr$.}
\vspace{0.3cm}
\label{f:TNGtests}
\end{figure*}

\Fig{TNGtests} is a combination analogous to \Figs{DMtests_Vmax}{DMtests_distancemetric}, but for test simulations based on the TNG model rather than on DM-only. \Fig{TNGtests_Vmax} presents $V_{c,{\rm max}}$ differences and \Fig{TNGtests_distancemetric} RMS particle position differences, in both cases the left panel is on a logarithmic time axis and the right panel on a linear time axis limited to $t\leq20\Myr$. First we note that in the fiducial case (orange squares) we do not see a gradual growth of the $V_{c,{\rm max}}$ differences as in \Fig{DMtests_Vmax} but they appear directly at a level of $\gtrsim10^{-4}\dex$. The same holds even for the case with shorter time steps (magenta triangles), where earlier snapshots are available, and where this level of differences is then seen as early as $2\times10^5\yr$ after the perturbation. Also, the RMS particle position differences show a power-law growth with time as early as can be probed, instead of the exponential growth seen in \Fig{DMtests_distancemetric}. This suggests that a different mechanism is at play in generating the differences with respect to the DM-only case.

We can gain significant insight from an additional set that has no analog in the DM-only case, in which the treatment of random numbers is modified compared to the fiducial model (and at the same time the maximum time step is limited in the same way as discussed above, in order to have high accuracy and high time resolution). In the fiducial TNG model\footnote{Note that this aspect of the TNG model is inherited from the original star-formation subgrid model of \citet{SpringelV_03a}.}, there is a single stream of random numbers that is used on each MPI task during the calculation, and each time a star-forming gas cell requires a random number to determine whether to turn into a star or wind particle, it draws the next number from that stream. This means that changes such as to the number of star-forming cells in the simulation, the way they are distributed between MPI tasks, or the sizes of their individual time steps, all necessarily affect the random number series that {\it each and every} gas cell in the simulation is using. This in turn means that, for example, changing one cell in the simulation box from non-star-forming to star-forming will immediately result in star formation and wind ejection events occurring in modified positions and times throughout the whole simulation volume (and hence, clearly, involving a superluminal flow of information).

In an attempt to control and mitigate this unphysical effect, which normally has no adverse consequences but critically pertains to our study here, we introduce a change where the random numbers are a deterministic function of time step and of coarse-grained position (`method 1' in Table \ref{t:verification}). In other words, for each simulation time step and coarse-grained spatial position, there exists a single, well-defined random number. The level of coarse-graining used is $1\hkpc$. In this way, a change to a single cell as in the example above only propagates through more `local' effects. For example, a change in one cell will affect the dynamics of its neighboring cells, some of which then might move from one coarse-grained voxel to the other and hence have their random number series changed, inducing further changes around them. Such cascade is, however, expected to take longer to develop than in the fiducial case. However, once the changes are significant enough to induce a change in the overall time stepping sequence of the simulation, by skipping even a single time step in one of the simulations with respect to the other, due to even a single particle requiring a shorter time step, then in subsequent times the behavior will be identical to the fiducial case, in that every cell in the simulation will be affected by a different set of random numbers between the two shadow simulations.

The results from the set with modified random number treatment according to this `method 1' are presented in \Fig{TNGtests} with purple pentagrams. In \Fig{TNGtests_Vmax} we indeed find that initially the $V_{c,{\rm max}}$ differences are at a very small level of $\sim10^{-8}\dex$, similarly to the very early times in the DM-only case. They then show a step function to the $\sim10^{-4}\dex$ level of the fiducial case at $t\approx7\Myr$. We interpret this delay to imply that for the first $\approx7\Myr$ after the perturbation, the random number sequences that determine the evolution of individual cells are for the most part identical between the two shadow simulations. This is confirmed by examining the time step sequences of the two, which indeed are found to be identical in this case for $156$ steps representing $7.1\Myr$ of evolution, at which point one of the two simulations introduces one additional short time step as required for evolving one single cell at an earlier time than its shadow simulation, thereby decoupling the random number sequences of the two simulations from each other.

The decoupling of the random number sequences between the two shadow simulations has a strong effect on the evolution of the RMS particle position differences, seen in \Fig{TNGtests_distancemetric}. Following that critical time, which in the fiducial case occurs essentially immediately at the perturbation time, and in the modified case occurs at $t=7.1\Myr$, the RMS distance between shadow particles evolves very accurately like a power law with index $2.5$ for approximately $2\Gyr$. We do not yet have an explanation for this behavior. Note that in the case of the set with modified random numbers, this quantity goes as $(t-7.1\Myr)^{2.5}$, which accounts for the steep transition region seen in the left panel around $7\times10^6\yr$.

In a further numerical experiment (`method 2' in Table \ref{t:verification}), we make the random numbers a deterministic function of both coarse-grained space and coarse-grained cosmological scale factor, such that they are independent of the nuisance parameter that is the sequential time step number of the simulation. In this case, the transition to a regime where all cells see different random number sequences between the two simulations occurs more gradually. This can be seen in all panels of \Fig{TNGtests} (gray crosses): the growth of the RMS distance (\Fig{TNGtests_distancemetric}) does not show as sharp a transition as does `method 1' at $7.1\Myr$. This is because the number of cells that develop different coarse-grained spatio-temporal evolution tracks grows gradually, with only local influences between cells\footnote{The shorter maximum time step we applied in `method 2' compared to `method 1' accounts for the slower initial growth, analogously to the DM-only case discussed in Appendix \ref{s:verification_DM}.}. This results also in a more gradual evolution of $V_{c,{\rm max}}$ differences (\Fig{TNGtests_Vmax}), which in this case take almost $10\Myr$ to grow from $\sim10^{-8}\dex$ to $10^{-4}\dex$ (compared to $1\Myr$ with `method 1'). The typical timescale of the evolution of the RMS distances at early times after the perturbation is a few million years, as indicated in the right panel of \Fig{TNGtests_Vmax}. This timescale is $\sim20$ times shorter than the $\sim117\Myr$ found in the DM-only case with the same small time steps, corresponding to the shorter dynamical times of galaxies compared to dark matter halos.

Finally, we present a test that is identical to the simulations introduced in Section \ref{s:no_random}, which completely remove the usage of random numbers, except for a much shorter time step (dark blue curves in \Fig{TNGtests}). These show an evolution of the differences, both in particle distances and in $V_{c,{\rm max}}$, that resembles an exponential growth with a timescale of $\approx4-10\Myr$, as indicated in the right panels. These simulations do not show the same rapid convergence toward the fiducial case that is seen in the other tests once the random numbers go out of sync between the shadow simulations in a pair. Instead, the evolution transitions directly from the exponential growth regime (on a timescale comparable to the dynamical timescale of galaxies) into the late power-law regime when the $V_{c,{\rm max}}$ differences become of order $10^{-5}\dex$.

We conclude that the use of random numbers in our simulations (and possibly by extension cosmological simulations produced by other codes) injects instantaneously a relatively high level of differences between shadow simulations, levels that may take of order billions of years to develop under exponential growth where there is no usage of random numbers. While the particular workings of these random numbers is clearly a numerical construct, it is an interesting -- and pertinent -- question whether they have analogs in the real universe. This is discussed in Section \ref{s:summary}. It is also worth pointing out that once such differences appear, they continue developing more slowly over cosmic time, as described in the main part of the paper. In this sense, the final $z=0$ differences are only seeded by the random number differences, but not directly determined by them, and indeed they generally develop to very different levels depending on whether feedback is present or not.

Hence, we identify three regimes to the development of the initial perturbations. First is the `chaotic' regime, which is similar to the DM-only case, in which the growth of perturbation is exponential. Second is the `injection' regime, in which the very small perturbations are very rapidly blown up by the injection of randomness into the star-formation process through the random numbers. This phase does not exist in the DM-only case or when random numbers are not used in the subgrid models. Third is the slower `dynamical' regime, during which the perturbations continue growing as a power-law or slower, in some cases reaching a plateau after several $\Gyr$. They typically grow into percent-level or even larger differences, which often constitute a sizable fraction of the overall variation within the galaxy population.

\section{(In)sensitivity to the group finder}
\label{s:FOFresults}

To mitigate a potential concern that the results we present in the main text are significantly affected by properties of the \SUBFIND{ }group structure algorithm, here we present results that are not based on \SUBFIND{ }in any way. Such a concern may arise because \SUBFIND{ }is not fully translationally and rotationally invariant. It may return different results in response to small changes in particle coordinates due to its use of a tree structure, and for the same reason its results are also generally not completely invariant to certain nuisance parameters. By calculating quantities only based on the raw particle data and on the Friends-Of-Friends algorithm, as we do in this appendix, we verify that these properties of \SUBFIND{ }do not, in fact, affect our results in any significant way.

\begin{figure*}
\centering
\includegraphics[width=1.0\textwidth]{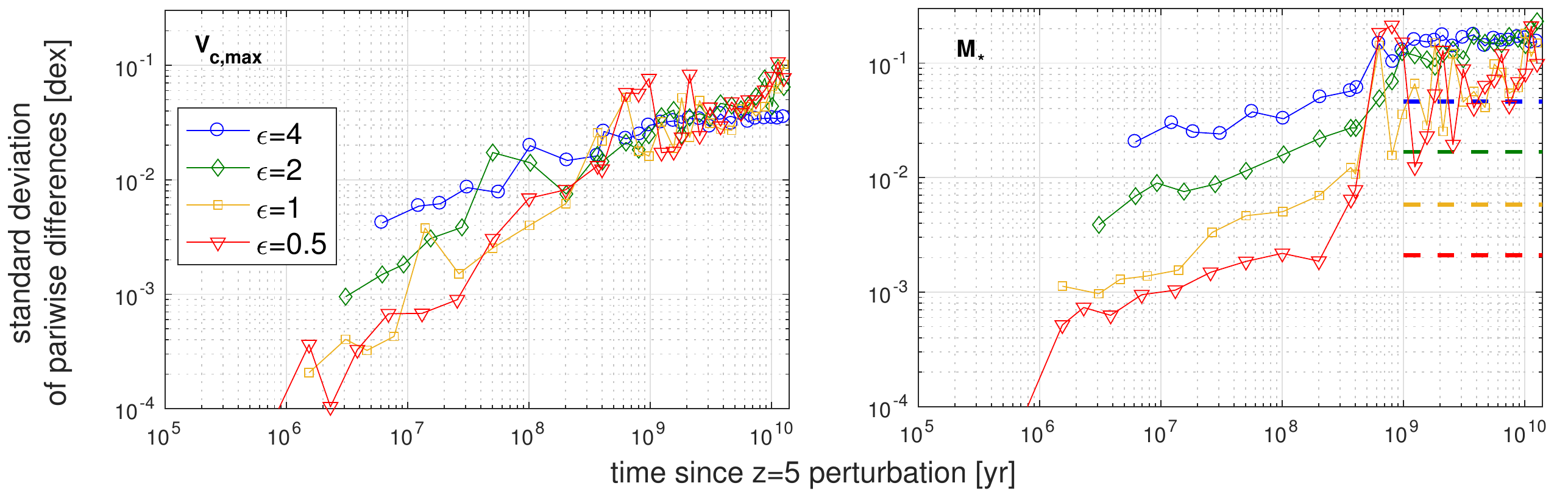}
\caption{The evolution of pairwise differences between shadow galaxies with final mass of $9.5<\log M_*[h^{-1}\Msun]<10$ in our TNG model simulation series, similarly to \Fig{differences_vs_time_TNG}, only when the properties are not based on \SUBFIND. The results are, however, very similar.}
\vspace{0.3cm}
\label{f:differences_vs_time_TNG_FOF}
\end{figure*}

\Fig{differences_vs_time_TNG_FOF} is analogous to the top row in \Fig{differences_vs_time_TNG}, and shows the growth over time of pairwise differences between the maximum circular velocities and the masses of shadow galaxies in our TNG model series. Here, however, these two quantities are calculated in a different way from that used for \Fig{differences_vs_time_TNG}, avoiding the use of \SUBFIND. The mass (right panel) is simply the full stellar mass assigned to Friends-Of-Friends halos. The maximum circular velocity (left panel) is calculated within a fixed aperture of $10\hkpc$ around the stellar particle with the lowest potential energy.

Quantitatively, the results in \Fig{differences_vs_time_TNG_FOF} are similar to those in \Fig{differences_vs_time_TNG}. In fact, the difference between the two is such that the pairwise differences are somewhat larger in the \SUBFIND-independent case shown here. We interpret that to be a result of the difference aperture that is used here (which itself is chosen so as to avoid the use of \SUBFIND), rather than a direct consequence of the group finding algorithm. This also means that we believe that the results presented in the main body of the paper are not driven or dominated by \SUBFIND.

It is worth pointing out, however, a general artifact of group finding, not specifically of \SUBFIND, that occurs in rare cases. This is where the timing of a merger is different between two shadow simulations such that in one shadow simulation there are two nearby galaxies that are still considered separate objects while in the other they are already considered as merged. In such a situation, the galaxy properties calculated by the group finder may even be different on the order of unity between the two shadows, while the state of the physical system itself is almost identical. In these cases, the large pairwise difference is an outlier to the Gaussian-like distribution of pairwise differences such as those shown in \Fig{1d_hists}, but it can affect the overall standard deviation of the distribution. An outcome of this can be seen most prominently in the $R_{1/2,*}$ and $M_*$ panels of \Fig{differences_vs_time_SP} at $t=5\times10^7\yr$, where the outlier point of the $\epsilon=0.5$ level is affected by a single galaxy, which in two of the shadows has just `absorbed' a satellite while in the other two has not yet. In such highly non-Gaussian cases, the simplistic estimate for the error on this standard deviation (shown as error bars for a few examples in \Fig{differences_vs_time_SP}) dramatically underestimates the true one, as expected.

\listofchanges

\end{document}